\documentclass[aps,prd,preprint,superscriptaddress,showpacs,floatfix,preprintnumbers,nofootinbib]{revtex4-1}

\usepackage[utf8]{inputenc}

\usepackage{cancel}

\usepackage[
    letterpaper, 
    left = 2.5cm, 
    right = 2.5cm, 
    top = 2.4cm, 
    bottom = 2.4cm, 
    headsep = 0.5cm, 
    footskip = 1.5cm]{geometry}

\usepackage[dvipsnames]{xcolor}      
\definecolor{lcolor}{rgb}{0.5,0,0}
\definecolor{citcolor}{rgb}{0,0.0,1}

\usepackage[breaklinks,colorlinks,urlcolor=blue,citecolor=citcolor,linkcolor=lcolor,linktoc=all]{hyperref}
\usepackage{color}
\usepackage{graphicx}	
\usepackage[export]{adjustbox}
\graphicspath{{./figures/}}
\usepackage[utf8]{inputenc}
\usepackage{amsmath} 
\usepackage{amssymb}
\usepackage[capitalise]{cleveref}

\usepackage[ragged]{footmisc}
\usepackage{slashed} 
\usepackage{float}

\makeatletter
\g@addto@macro\bfseries{\boldmath}
\makeatother

\usepackage{tikz}
\usepackage[customcolors]{hf-tikz}
\usepackage{mciteplus}

\usetikzlibrary{arrows,cd,shapes,decorations.pathmorphing,decorations.markings,shadings}
\tikzset{
  big arrow/.style={
    decoration={markings,mark=at position 1 with {\arrow[scale=4,#1]{>}}},
    postaction={decorate},
    shorten >=0.4pt},
  big arrow/.default=blue}

\newcommand{\Ac}{\mathcal{A}}
\newcommand{\Act}{\tilde{\mathcal{A}}}

\newcommand{\bbone}{\text{\usefont{U}{bbold}{m}{n}1}}
\MakeRobust{\bbone}

\newcommand{\Lbar}{\overline{\Lambda}}
\newcommand{\mE}{m_{\text{E}}}

\newcommand{\ca}{C_A}
\newcommand{\nc}{N_c}
\newcommand{\cf}{C_F}
\newcommand{\da}{d_A}
\newcommand{\nf}{N_f}

\newcommand{\tf}{T_F}
\newcommand{\MSbar}{$\overline{\mathrm{MS}}$}

\newcommand{\pt}{{\vec{p}}}

\newcommand{\vt}{{\vec{v}}}

\renewcommand{\vec}{\mathbf}

\newcommand{\Afunc}{A}
\newcommand{\Bfunc}{B}

\renewcommand{\epsilon}{\varepsilon}

\newcommand{\ud}{\mathrm{d}}

\newcommand{\gs}{g_s}

\newcommand{\gamE}{{\gamma_{\text{E}}}}
\newcommand{\tr}[1]{\operatorname{Tr}\left[ #1 \right] }

\begin{document}

\title{Soft gluon self-energy at finite temperature and density: hard NLO corrections in general covariant gauge}

\preprint{HIP-2023-9/TH,~TUM-EFT 180/23}
\author{Tyler Gorda}
\email{tyler.gorda@physik.tu-darmstadt.de}
\affiliation{Technische Universit\"{a}t Darmstadt, Department of Physics, 64289 Darmstadt, Germany}
\affiliation{ExtreMe Matter Institute EMMI, GSI Helmholtzzentrum f\"ur Schwerionenforschung GmbH, 64291 Darmstadt, Germany}
\author{Risto Paatelainen}
\email{risto.paatelainen@helsinki.fi}
\affiliation{Department of Physics and Helsinki Institute of Physics, P.O.~Box 64, FI-00014 University of Helsinki, Finland}
\author{Saga S\"appi}
\email{saga.saeppi@tum.de}
\affiliation{TUM Physik-Department, Technische Universität München, James-Franck-Str. 1, 85748 Garching, Germany}
\affiliation{Excellence Cluster ORIGINS, Boltzmannstrasse 2, 85748 Garching, Germany}
\author{Kaapo Seppänen}
\email{kaapo.seppanen@helsinki.fi}
\affiliation{Department of Physics and Helsinki Institute of Physics, P.O.~Box 64, FI-00014 University of Helsinki, Finland}

\begin{abstract}

We compute the next-to-leading order (NLO) hard correction to the gluon self-energy tensor with arbitrary soft momenta in a hot and/or dense weakly coupled plasma in Quantum Chromodynamics. Our diagrammatic computations of the two-loop and power corrections are performed within the hard-thermal-loop (HTL) framework and in general covariant gauge, using the real-time formalism. We find that after renormalization our individual results are finite and gauge-dependent, and they reproduce previously computed results in Quantum Electrodynamics in the appropriate limit. Combining our results, we also recover a formerly known gauge-independent matching coefficient and associated screening mass in a specific kinematic limit. Our NLO results supersede leading-order HTL results from the 1980s and pave the way to an improved understanding of the bulk properties of deconfined matter, such as the equation of state.
\end{abstract}

\maketitle

\tableofcontents

\newpage
\section{Introduction}
\label{sec:intro}

In a weakly coupled quark-gluon plasma (QGP) at large temperature $T$ or chemical potential $\mu$, the propagation of massless gluonic modes is qualitatively modified through interactions with medium fluctuations, leading to the phenomena of dynamical screening. Such screening primarily affects long-wavelength gluonic modes, with energies of the order of $\gs T$ or $\gs \mu$, where $\gs$ is the strong gauge coupling. For these ``soft'', long-wavelength gluons, interactions with the highest-energy partons within the QGP (with energy proportional to $T$ or $\mu$, dubbed a ``hard'' energy scale) qualitatively modify their dispersion relation, signalling the emergence of a thermal mass scale.

Due to the large separation of scales $\gs T \ll T$ in a weakly coupled QGP, the dynamics of long-wavelength gluons can be studied within an effective field theory known as Hard-Thermal-Loop (HTL) effective field theory \cite{Braaten:1989mz} (or Hard-Dense-Loop in the specific case of large $\mu$ and small $T$ \cite{Ipp:2003cj,Ipp:2006ij}). This name arises from considering the interactions of soft gluons within the language of Feynman diagrams; in this language, soft external gluon propagation (or interactions between multiple soft gluons) becomes corrected through interactions with hard internal loop momenta.
Including these additional HTL propagators and vertices in higher-order diagrams contributing to soft scattering processes or thermodynamics protects the calculations from infrared (IR) divergences that are otherwise present (see e.g.~\cite{Ghiglieri:2020dpq} for a review). The HTL framework has been successfully applied to a number of problems. For example, in the context of high-temperature Quantum Chromodynamics (QCD) the static gluon damping rate and the nonabelian Debye screening mass were first computed to leading order in $\gs$ in \cite{Braaten:1990it} and \cite{Kajantie:1981hu}, respectively. The computation of the Debye mass was later generalized to next-to-leading order (NLO) in \cite{Rebhan:1993az,Arnold:1995bh}. In addition, there are several dynamical quantities that are sensitive to the soft $\gs T$ scale, such as thermal photon production rates \cite{Aurenche:1998nw,Arnold:2002ja,Ghiglieri:2013gia,Jackson:2022fqj}, jet and heavy quark energy loss \cite{Braaten:1991jj,Braaten:1991we,Baier:1996kr}, nonrelativistic heavy quark diffusion \cite{Caron-Huot:2007rwy,Caron-Huot:2008dyw} and transport coefficients \cite{Arnold:2003zc,Ghiglieri:2018dib,Danhoni:2022xmt}. In the context of $T = 0$ and large $\mu$, the HTL framework has been a crucial part of recent advances in the evaluation of the QCD pressure to next-to-next-to-next-to leading order (N3LO) \cite{Gorda:2021znl,Gorda:2021kme}. 

Until recently, HTL has been nearly exclusively used in the sense which we refer to here as ``one-loop HTL''.\footnote{Note that this is distinct from the concept of e.g. the loop counting in HTLpt \cite{Haque:2014rua}.} That is to say, the HTL propagators and vertices are computed as soft limits of one-loop quantities. Corrections to one-loop HTL fall into two categories: The first is further loop corrections, and the second is what is referred to as power corrections. The former arises in the usual diagrammatic expansion, while the latter is related to higher-order expansions in the external gluonic momenta within the lower-order diagrams. 

In the context of Quantum Electrodynamics, both of the HTL corrections have been computed at both high $T$ \cite{Manuel:2016wqs,Carignano:2017ovz,Carignano:2019ofj} and high $\mu$ and arbitrary $T$ \cite{Gorda:2022fci,Gorda:2022zyc} using diagrammatic machinery and the real-time formalism, and they have recently been extended to the case of general gauge theories at high $T$ using a kinetic-theory approach in Feynman gauge \cite{Ekstedt:2023anj}. In this work, we compute both the two-loop HTL and power corrections to the gluon self-energy for large $T$ and $\mu$ in general covariant gauge within a diagrammatic approach using the real-time formalism.

Note that in order to compute the full self-energy, and infer from it physical properties of soft gluons, such as dispersion relations, one must compute not only these higher-order corrections to the HTL theory, but one must also calculate diagrams \emph{using} the HTL theory. That is, to calculate the gluon dispersion relation at high $T$ and/or $\mu$ to $O(\gs^4)$, one must also compute one-loop diagrams with soft, HTL-resummed internal gluon lines. Such corrections in fact dominate over the two-loop and power corrections calculated in this work at high $T$ \cite{Mirza:2013ula}: Due to the Bose enhancement, they get lifted to $O(g_s^3)$, and even the resummed \emph{two-loop} diagrams contribute to the full $O(g_s^4)$ self-energy, being lifted from $O(g_s^6)$ for the same reason. Even at small $T$, the one-loop resummed diagrams still compete with the corrections to HTL at $O(g_s^4)$. We emphasise that, unless otherwise stated, in the terminology used in this work after this section, `NLO' always refers to next-to-leading order in the HTL expansion,  which will be discussed in more detail in \cref{subsec:HTLlimit}, instead of the complete self-energy, for which the calculation presented here would constitute only a partial NLO contribution at small $T$ and a partial next-to-next-to-leading order (NNLO) contribution at large $T$. We will not consider the soft resummed diagrams in our computation here, leaving them for future work. We do however note that the corrections to HTL bring valuable information on their own. For example, within the context of the N3LO pressure of QCD at $T = 0$, the effect of these corrections has already been computed in \cite{Gorda:2021znl, Gorda:2021kme}, and a distinct missing contribution can be attributed to the NLO corrections to HTL alone. Diagrammatically, the two types of corrections to the self-energy contribute to the pressure in a rather distinct way, as the resummed diagrams contribute through two-loop HTL topologies, while the terms computed in this paper contribute through insertions on the one-loop HTL ring sum.

There is one main subtlety related to the evaluation of the two-loop self-energy in general covariant gauge that is not present in the standard one-loop case and that we will encounter in this work: namely, the tensor structure. Generally, the Ward identities of QCD are weaker than those in QED and do not ensure transversality of the self-energy $\Pi^{\mu \nu}(K)$ with respect to the external four-momentum $K$ \cite{Weldon:1996kb}. In the one-loop HTL case, both transversality and gauge independence do hold (both of which are related due to the Ward identities), and the basis of rank-two tensors is two-dimensional (taking into account the symmetry as well as the remnant $\mathrm{SO}(d)$-invariance, with $d$ the number of spatial dimensions). At higher orders, this is not guaranteed.
Indeed, even transversality of the self-energy with respect to $K^\mu K^\nu$ is broken at $O(g_s^4)$ for general external momenta.
As we will see here, it also turns out that both transversality and gauge independence are broken individually for the two-loop HTL and power correction contributions to the self-energy, although they may be restored for specific combinations and kinematic limits of these two components. For general kinematics, these properties are expected to break down further. 

The organization of this work is as follows. We begin in \cref{sec:setup} with a brief overview of the setup and details, conventions, and a summary of the real-time formalism. In \cref{sec:gluon_se_structure}, we present the general structure of the gluonic self-energy and a general formalism for the HTL expansion of the quantity. We then proceed in \cref{sec:organization} to give details of the calculation, with our Results and Discussion presented in \cref{sec:summaryoftheres}. %
Many technical details of the calculations are explained in the Appendices.

\section{Setup}
\label{sec:setup}

\subsection{Conventions}
\label{sec:conventions}

Our conventions and notation are as follows. We work in $D= 4-2\varepsilon$ spacetime dimensions and $d = D-1$ spatial dimensions, and in Minkowskian space with the mostly plus Minkowskian metric $g_{\mu\nu} = \text{diag}(-1,+1,+1,+1)$. We write the components of four-vectors as
\begin{equation}
P \equiv (p^0,\mathbf{p}), \quad p \equiv \vert \mathbf{p}\vert,
\end{equation}
with $P^2 = -p_0^2 + p^2$, where the individual spatial components are $p^i$ with $i=1,\dots,d$. We regulate the divergent parts of the loop diagrams within dimensional regularization. The $D$-dimensional integration measure is defined as
\begin{equation}
\int_{P} \equiv \left (\frac{e^{\gamE}\Lbar^2}{4\pi}\right )^{\frac{4-D}{2}} \int \frac{\mathrm{d}^D P}{(2 \pi)^D} = \int_{-\infty}^\infty \frac{\mathrm{d}p^0}{2 \pi} \int_\mathbf{p} \,,
\end{equation}
where the shorthand $\int_\mathbf{p}$ denotes the spatial part of the integration 
\begin{equation}
\int_\mathbf{p}  \equiv \left (\frac{e^{\gamE}\Lbar^2}{4\pi}\right )^{\frac{3-d}{2}} \int \frac{\mathrm{d}^d \mathbf{p}}{(2 \pi)^d}.
\label{eq:intmeasure}
\end{equation}
Here, the factor 
$(e^{\gamE}/4\pi)^{(3-d)/2}$, with $\gamE$ the Euler--Mascheroni constant, is introduced as usual to simplify the final expressions, and $\Lbar$ is the $\overline{\text{MS}}$  renormalization scale.

The SU$(\nc)$ group theory factors that appear in the calculation are given by: 
\begin{equation}
\begin{split}
\ca \delta^{cd} & = f^{abc} f^{abd} = \nc \delta^{cd},\\
\da & = \delta^{aa} = \nc^2 -1,\\
\cf \delta_{ij} & = (T^a T^a)_{ij}  = \frac{\nc^2 -1}{2\nc} \delta_{ij}.
\end{split}
\end{equation}
Here, $\nc$ is the number of colors, $f^{abc}$ is totally anti-symmetric with respect to the interchange of any pair of its indices and the generators $T^a$ in the fundamental representation are normalized according to $\tr{T^a T^b} =  \delta^{ab}\tf$ with $\tf = 1/2$. In addition, the fermion loop comes with the factor of $\nf$, where $\nf$ is the number of fermions.

\subsection{Real-time formulation of thermal QCD}\label{subsec:realtimeQCD}

\begin{figure}
\begin{center}
\includegraphics[scale=1.0]{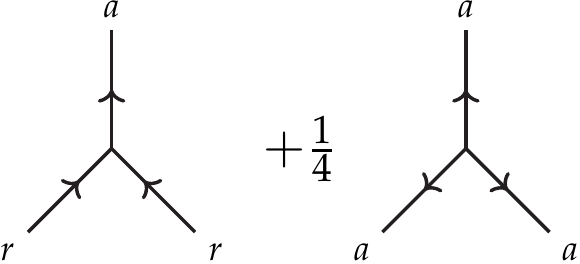}
$\qquad\qquad$ %
\includegraphics[scale=1.1]{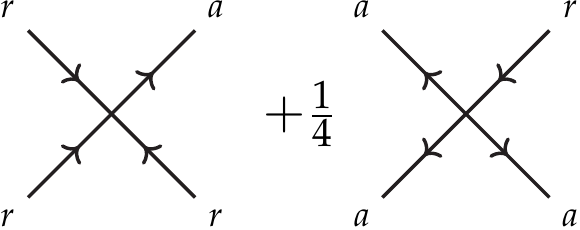}
\end{center}
\caption{%
Possible $r/a$ assignments for the (left) three-point and (right) four-point vertices in the real-time formalism. Vertices with three $a$ assignments receive an extra factor of $+1/4$.
\label{fig:raVertices}
}
\end{figure}

We use the real-time formulation of thermal QCD at finite temperature and density in the $r/a$ basis (for a recent review see e.g.~\cite{Ghiglieri:2020dpq}). We work in the $R_\xi$-class of (covariant) renormalizable gauges, where the free (or bare) retarded/advanced   $[D^{R/A}_{\mu\nu}]^{ab} \equiv  \delta^{ab}D^{R/A}_{\mu\nu}$  and the
symmetric $[D^{rr}_{\mu\nu}]^{ab}  \equiv \delta^{ab}D^{rr}_{\mu\nu}$ gluon propagators are given by 
\begin{equation}
\begin{split}
D^{R}_{\mu\nu}(P) & = \includegraphics[valign=c,scale=1.2]{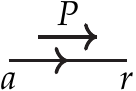} \equiv g_{\mu\nu}\Delta^{R}(P) - i(1-\xi) P_\mu P_\nu [\Delta^{R}(P)]^2, \\
D^{A}_{\mu\nu}(P) & = \includegraphics[valign=c,scale=1.2]{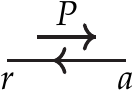} \equiv g_{\mu\nu}\Delta^{A}(P) - i(1-\xi) P_\mu P_\nu [\Delta^{A}(P)]^2, \\
D_{\mu\nu}^{rr}(P) & = \includegraphics[valign=c,scale=1.2]{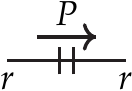} \equiv g_{\mu\nu} \Delta^{rr}_{B,1}(P) - i(1-\xi) P_\mu P_\nu \Delta^{rr}_{B,2}(P), 
\end{split}
\end{equation}
where $\xi$  is the gauge-fixing parameter, $\Delta^{rr}_{B,n}$ is the symmetric scalar propagator for bosons (defined below) and $\Delta^{R/A}$ is given by
\begin{equation}
\Delta^{R/A}(P) \equiv \frac{-i}{P^2 \mp i\eta p^0}.
\end{equation}
Here, $\eta>0$ determines the causal structure of the propagator and will either drop out or in the end be taken to zero for physical quantities. For the ghosts, the retarded/advanced $[\tilde{D}^{R/A}]^{ab} \equiv \delta^{ab}\tilde{D}^{R/A}$ and symmetric $[\tilde{D}^{rr}]^{ab} \equiv \delta^{ab}\tilde{D}^{rr}$ propagators are simply
\begin{equation}
\tilde{D}^{R/A}(P) \equiv \Delta^{R/A}(P), \quad\tilde{D}^{rr}(P) \equiv \Delta^{rr}_{B,1}(P), 
\end{equation}
and for the fermions, the retarded/advanced $[S^{R/A}]_{ij} \equiv \delta_{ij}S^{R/A}$ and symmetric $[S^{rr}]_{ij} \equiv \delta_{ij}S^{rr}$ propagators read
\begin{equation}
S^{R/A}(P) \equiv  -\slashed{P} \Delta^{R/A}(P), \quad S^{rr}(P) \equiv -\slashed{P} \Delta^{rr}_F(P).
\end{equation}
The symmetric scalar propagators $\Delta^{rr}_{B,n}$ and $\Delta^{rr}_F$ are related to the retarded and advanced ones through the Kubo--Martin--Schwinger (KMS) relation
\begin{equation}
\label{eq:KMSrelation}
\Delta^{rr}_{B,n}(P) \equiv N_B(P)\Delta^d_n(P), \quad \Delta^{rr}_F(P) \equiv N^{-}_F(P) \Delta^d_1(P),
\end{equation}
where the functions $N_B$ and $N_F$ are given in terms of the bosonic and fermionic distribution functions 
\begin{equation}
N_B(P) \equiv \frac{1}{2} + n_B(p^0), \quad N_F^{\pm}(P) \equiv \frac{1}{2}- n_F(p^0 \pm \mu),
\end{equation}
with $n_{B/F}(p^0) \equiv (e^{p^0/T} \mp 1)^{-1}$. The function $\Delta^d_n$ is defined as the difference of the $n$th powers of retarded and advanced propagators
\begin{equation}
    \Delta_n^d(P) \equiv \Delta^R(P)^n - \Delta^A(P)^n,
\end{equation}
where we usually abbreviate $\Delta^d\equiv\Delta_1^d$. In our calculations, we utilize the following formula originating from the residue theorem \cite{Gorda:2022fci}
\begin{equation}\label{eq:Deltadresidue}
\int_\mathbb{R} \frac{\mathrm{d}p^0}{2\pi} \Delta_n^d(P) f(p^0) = (-i)^{n+1} \sum_\pm \mathrm{Res}\left[ \frac{f(p^0)}{(P^2)^n},p^0=\pm p \right],
\end{equation}
where $f$ is not singular at $P^2=0$. In the case $n=1$, the above equation corresponds to the familiar Sokhotski–Plemelj formula,
\begin{equation}\label{eq:scalarspectralfunc}
\Delta^d_1(P) = \frac{-i}{P^2 - i\eta p^0}-\frac{-i}{P^2 + i\eta p^0} = 2\pi \, \mathrm{sgn}(p^0) \delta(P^2) = \frac{\pi}{p} \big(\delta(p-p^0)-\delta(p+p^0)\big) \,.
\end{equation}
Further, we frequently rely on the parity properties of the functions above,
\begin{equation}\label{eq:funcparity}
\begin{alignedat}{3}
\Delta^A(P) &= \Delta^R(-P) \,, &\qquad \Delta^d_n(P) &= -\Delta^d_n(-P) \,, \\
  N_B^{ }(P) &=-N_B^{ }(-P) \,, &\qquad N_F^\pm(P) &= -N_F^\mp(-P) \,.
\end{alignedat}
\end{equation}
In addition to propagators, we also need all the possible three- and four-point  QCD vertices coming from the gluon self-interactions and interactions between the fermion and gluon. In the $r/a$ basis, there are two distinct ways to assign $r/a$ labels to three- and four-point vertices. These are $rra$, $rrra$, $aaa$, and $aaar$, in which the latter two vertices are multiplied with an extra factor of $+1/4$. These vertices are drawn in \cref{fig:raVertices} by using the same graphical causal arrow representation we introduced for the propagators.

Finally, we note that the complete set of QCD Feynman rules for the propagators and vertices in general covariant gauges are given in \cref{sec:appA} of this paper.

\section{Structure of the gluon self-energy}
\label{sec:gluon_se_structure}

In the real-time formalism, the gluon self-energy (or amputated two-point function) becomes a $2\times 2$ matrix 
\begin{equation}
\mathbf{\Pi} = 
\begin{pmatrix}
0 & \Pi^A \\
\Pi^R & \Pi^{aa}
\end{pmatrix}.
\end{equation}
The retarded/advanced self-energy tensor $[\Pi^{R/A}_{\mu\nu}]^{ab} \equiv \delta^{ab} \Pi^{R/A}_{\mu\nu}$ of the gluon field is defined through the Dyson--Schwinger equation as 
\begin{equation}
\label{eq:dysoneq}
i\Pi^{R/A}_{\mu\nu}(K) = (\hat{D}^{R/A}_{\mu\nu})^{-1}(K) - (D^{R/A}_{\mu\nu})^{-1}(K),
\end{equation}
where $(\hat{D}^{R/A}_{\mu\nu})^{-1}$ and $(D^{R/A}_{\mu\nu})^{-1}$ are the inverse full and free gluon propagators, respectively. The form of the gluon self-energy tensor is further constrained by requiring gauge-invariance of various QCD Green's functions resulting in the Slavnov--Taylor identities \cite{Slavnov:1972fg,Taylor:1971ff} (non-Abelian generalization of Ward--Takahashi identities). For example, in the $R_\xi$-class of (covariant) renormalizable gauges, the full retarded gluon propagator satisfies the following identity \cite{Weldon:1996kb},
\begin{equation}
K^\mu K^\nu \hat{D}^R_{\mu\nu}(K) = -i \xi.\label{eq:STidentity}
\end{equation} This relation constrains the self-energy tensor through the Dyson--Schwinger equation in \cref{eq:dysoneq}. It is important to note that the Slavnov--Taylor identities do not fix the self-energy itself to be gauge-invariant: The dependence on $\xi$ is seen to disappear quite easily for the leading-order HTL self-energy, but generally, only physical quantities such as screening masses must be gauge-independent. 

\subsection{Spacetime rank two tensor basis}

In the vacuum, the only available tensor structures for $\Pi_{\mu\nu}$ are $g_{\mu\nu}$ and $K_\mu K_\nu$, owing to Lorentz symmetry and the symmetry of the self-energy in the Lorentz indices. The application of \cref{eq:STidentity} in that case requires the self-energy to be transverse to its momentum,
\begin{equation}
K^\mu \Pi_{\mu\nu}(K) = 0 .
\end{equation}
Hence, the vacuum self-energy may be written as
\begin{equation}
\Pi_{\mu\nu} (K) = \left(g_{\mu\nu}-\frac{K_\mu K_\nu}{K^2} \right)\Pi(K^2) \equiv \mathbb{P}_{\mu\nu}(K) \Pi(K^2) ,
\end{equation}
where $\mathbb{P}$ is a projector transverse $D$-dimensionally transverse to its argument, satisfying the usual property $\mathbb{P}_{\mu\lambda} {\mathbb{P}^\lambda}_\nu = \mathbb{P}_{\mu\nu}$ (idempotent), and $\Pi(K^2)$ is a Lorentz scalar.

In a thermal medium, the Lorentz symmetry is broken by the rest frame of the thermal bath $n^\mu$, but the symmetry in the Lorentz indices is maintained. Consequently, the tensor basis for the self-energy extends to four different tensors: $g_{\mu\nu}$, $K_\mu K_\nu$, $n_\mu n_\nu$ and $n_\mu K_\nu+K_\mu n_\nu$. For simplicity, we will choose to work in the rest frame of the thermal medium so the remaining symmetry is associated with spatial rotations, and the velocity of the medium satisfies $n^\mu = (1,\mathbf{0})$. 
When considering the transversality properties of the self-energy, it is convenient to define a vector $\tilde{n}_\mu (K) = \mathbb{P}_{\mu\nu}(K) n^\nu$, and choose the tensor basis as linear combinations of the above tensors:
\begin{align}
\label{eq:projectors_L}
\mathbb{P}^\mathrm{L}_{\mu\nu} &\equiv \frac{\tilde{n}_\mu \tilde{n}_\nu}{\tilde{n}^2} , \\
\label{eq:projectors_T}
\mathbb{P}^\mathrm{T}_{\mu\nu} &\equiv \mathbb{P}_{\mu\nu} - \mathbb{P}^\mathrm{L}_{\mu\nu} = \delta_\mu^i \delta_\nu^j \left(g_{ij}-\frac{k_i k_j}{k^2}\right) , \\
\label{eq:projectors_C}
\mathbb{P}^\mathrm{C}_{\mu\nu} &\equiv \frac{1}{k}\left(\tilde{n}_\mu K_\nu + K_\mu \tilde{n}_\nu \right) , \\
\label{eq:projectors_D}
\mathbb{P}^\mathrm{D}_{\mu\nu} &\equiv \frac{K_\mu K_\nu}{K^2} .
\end{align}
One may check that $\mathbb{P}^\mathrm{T}$, $\mathbb{P}^\mathrm{L}$ and $\mathbb{P}^\mathrm{D}$ are idempotent and mutually orthogonal. On the other hand, $\mathbb{P}^\mathrm{C}$ satisfies the relations
\begin{align}
\mathbb{P}^\mathrm{T}_{\mu \lambda} {\mathbb{P}^{\mathrm{C} \lambda}}_\nu &= 0 , \\
\mathbb{P}^\mathrm{C}_{\mu \lambda} {\mathbb{P}^{\mathrm{C} \lambda}}_\nu &= -\mathbb{P}^\mathrm{L}_{\mu\nu}-\mathbb{P}^\mathrm{D}_{\mu\nu} ,\\ \mathbb{P}^\mathrm{L}_{\mu \lambda} {\mathbb{P}^{\mathrm{C} \lambda}}_\nu &= \frac{\tilde{n}_\mu K_\nu}{k} , \\
\mathbb{P}^\mathrm{D}_{\mu \lambda} {\mathbb{P}^{\mathrm{C} \lambda}}_\nu &= \frac{K_\mu \tilde{n}_\nu}{k} ,
\end{align}
i.e., it is not idempotent and only orthogonal to $\mathbb{P}^\mathrm{T}$ but the contractions with $\mathbb{P}^\mathrm{L},\mathbb{P}^\mathrm{D}$ vanish once traced over, in particular implying $K^\mu K^\nu \mathbb{P}^\mathrm{C}_{\mu\nu} = 0$.
Further, it is easy to show that the projectors $\mathbb{P}^\mathrm{T}$ and $\mathbb{P}^\mathrm{L}$ are transverse to $K$.

By using the above projectors, the retarded gluon self-energy\footnote{From now on, every appearance to the self-energy $\Pi$ will implicitly refer to the retarded self-energy $\Pi^R$ unless otherwise specified.} decomposes as \cite{Weldon:1996kb}
\begin{equation}
\Pi^R_{\mu\nu} = \mathbb{P}^\mathrm{T}_{\mu \nu} \Pi_\mathrm{T} + \mathbb{P}^\mathrm{L}_{\mu \nu} \Pi_\mathrm{L} + \mathbb{P}^\mathrm{C}_{\mu \nu} \Pi_\mathrm{C} + \mathbb{P}^\mathrm{D}_{\mu \nu} \Pi_\mathrm{D} \label{eq:gluonselfcomp}
\end{equation}
in the rest frame of the thermal medium. The components now depend separately on $k^0$ and $k$ due to the broken Lorentz symmetry. By employing the above properties of the projectors, one can project out the individual components from the full tensor,
\begin{equation}
\begin{aligned}
\Pi_\mathrm{T} &= \frac{1}{d-1} \mathbb{P}^\mathrm{T}_{\mu\nu} \Pi^{\mu \nu} , &\qquad \Pi_\mathrm{L} &= \mathbb{P}^\mathrm{L}_{\mu\nu} \Pi^{\mu \nu} , \\
\Pi_\mathrm{C} &= -\frac{1}{2} \mathbb{P}^\mathrm{C}_{\mu\nu} \Pi^{\mu \nu} , & \Pi_\mathrm{D} &= \mathbb{P}^\mathrm{D}_{\mu\nu} \Pi^{\mu \nu} . \label{eq:gluonselfcompexpl}
\end{aligned}
\end{equation}

The full retarded gluon propagator can be obtained by inserting the free retarded propagators from the previous Section into the Dyson--Schwinger equation in \cref{eq:dysoneq} while imposing the Slavnov--Taylor identity in \cref{eq:STidentity}. This leads to a decomposition in terms of the four projectors
\begin{equation}
 \hat{D}_{\mu\nu}^R = \mathbb{P}^\mathrm{T}_{\mu\nu} \Delta_\mathrm{T} + \mathbb{P}^\mathrm{L}_{\mu\nu} \Delta_\mathrm{L} + \mathbb{P}^\mathrm{C}_{\mu\nu} \Delta_\mathrm{C} + \mathbb{P}^\mathrm{D}_{\mu\nu} \Delta_\mathrm{D} \,, \label{eq:fullgluonprop}
\end{equation}
where
\begin{align}
\Delta_\mathrm{T} &= \frac{-i}{K^2+\Pi_\mathrm{T}} ,\label{eq:gluonpropT} \\
\Delta_\mathrm{L} &= \frac{-i}{K^2+\Pi_\mathrm{L}}\left( 1 + \xi \frac{\Pi_D}{K^2}\right) , \label{eq:gluonpropL} \\
\Delta_\mathrm{C} &= \frac{-i}{K^2+\Pi_\mathrm{L}} \left( -\xi \frac{\Pi_\mathrm{C}}{K^2} \right) , \label{eq:gluonpropC} \\
\Delta_\mathrm{D} &= -\frac{i\xi}{K^2} , \label{eq:gluonpropD}
\end{align}
with the self-energy components satisfying the non-linear relation
\begin{equation} \label{eq:slavnovpiD}
\Pi_\mathrm{D} = -\frac{\Pi_\mathrm{C}^2}{K^2 + \Pi_\mathrm{L}} .
\end{equation}
Note that for notational simplicity, we have absorbed the $i\eta$ from the free propagator into $k^0$, i.e.~when the $i\eta$-prescription is relevant we must substitute $k^0 \rightarrow k^0+i\eta$. Dictated by the Slavnov--Taylor identity, the longitudinal component $\Delta_\mathrm{D}$ does not receive self-energy corrections and is determined by the bare propagator alone. Furthermore, contrary to the vacuum case, the Slavnov--Taylor identity does not require a transverse gluon self-energy, so generally $K^\mu \Pi_{\mu\nu} \neq 0$ in a thermal medium. The physical, propagating, degrees of freedom are still the transverse and longitudinal modes, and this is not changed by the newly non-vanishing self-energy components. The transverse propagator spans a two-dimensional subalgebra corresponding to two nontrivial modes, and the remaining three propagators, which are not linearly independent, span a two-dimensional subalgebra corresponding to the longitudinal mode and a non-propagating massless mode $K^2=0$ which is purely a gauge artifact. Together, these span the four-dimensional algebra of rank two symmetric tensors with one external scale at finite temperature.

As a self-consistency check, we have explicitly verified that within our computation ($R_\xi$ gauges with full kinematics) \cref{eq:slavnovpiD} holds perturbatively to the first non-trivial order in $g_s$, i.e.~at $O(g_s^4)$ we obtain $\Pi_\mathrm{D}^\text{2-loop}=-(\Pi_\mathrm{C}^\text{1-loop})^2/K^2$.
However, in the HTL limit it turns out that the self-energy component $\Pi_\mathrm{D}$ vanishes at order $O(g_s^4)$. Still, according to \cref{eq:slavnovpiD} it has to become nonzero at $O(g_s^6)$ for general values of $\xi$ due to the finite HTL contributions to $\Pi_\mathrm{C}$ and $\Pi_\mathrm{L}$ at $O(g_s^4)$ and $O(g_s^2)$ respectively, and the form of this term can already be determined.

\subsection{HTL limit}
\label{subsec:HTLlimit}

We discuss now the HTL limit in some generality. Consider an expansion of the gluon self-energy, with both vacuum and matter parts included, in small coupling and external momentum. In general, this expansion takes the form
\begin{equation}
\label{eq:Piexpand}
\Pi^{\mu\nu}(K) \equiv \sum_{j=1}^{\infty} \Pi_{j\text{-loop}}^{\mu\nu} \equiv \sum_{j=1}^{\infty} \sum_{p=0}^{\infty} \Pi_{j,p}^{\mu\nu} = \mE^2 \sum_{j=1}^{\infty}\sum_{p=0}^{\infty} \left (\frac{K^2}{\mE^2} \right )^p \gs^{2(j+p-1)} C^{\mu\nu}_{j,p}(\mu/T, k^0/k),   
\end{equation}
where the parameter $\mE^2$ is an $O(\gs^2T^2)$ effective thermal mass scale and the $C^{\mu\nu}_{j,p}$ are dimensionless functions. Here, the index $j$ represents the number of loops in a diagram with $j=1$ corresponding to the one-loop case, and the index $p$ represents the degree of power corrections, with $p=0$ corresponding to the strict HTL limit \cite{Braaten:1989mz}. For example, in a  notation adapted from \cite{Gorda:2022fci} for the first three:
\begin{equation}
\begin{split}
    \Pi_{1,\mathrm{HTL}}^{\mu\nu} & \equiv \Pi_{1,0}^{\mu\nu}  = \mE^2 C_{1,0}^{\mu\nu} \sim \mE^2, \\   
    \Pi_{2,\mathrm{HTL}}^{\mu\nu}  & \equiv \Pi_{2,0}^{\mu\nu}  =  \mE^2 \gs^{2} C_{2,0}^{\mu\nu} \sim \mE^2 \gs^2, \\
    \Pi_{1,\text{Pow}}^{\mu\nu} & \equiv \Pi_{1,1}^{\mu\nu}  = K^2 \gs^2 C_{1,1}^{\mu\nu} \sim K^2 \gs^2. \\
\end{split}
\label{eq:pi_notation}
\end{equation}
The latter two terms are what we focus on in this work.

In practice, computing the above terms in the HTL expansion of the gluon self-energy within a diagrammatic calculation can be implemented using the following steps: 

\begin{enumerate}
    \item The contributions to the self-energy $\Pi^{\mu \nu}$ is first written as a sum over diagrams $\mathcal{G}_i$.
    \item The contribution to $\Pi^{\mu\nu}$ from each single diagram $\mathcal{G}_i$ is contracted with the four basis tensors to obtain tensor components $\{ \Pi^{\mu\nu}_\mathrm{T}, \Pi^{\mu\nu}_\mathrm{L}, \Pi^{\mu\nu}_\mathrm{C}, \Pi^{\mu\nu}_\mathrm{D}\}_i$ for each diagram. %
    \item For each of the tensor components, the internal zero-component integrals are performed with the help of the $\Delta^d_n$-propagators, which set the loop momenta on shell [see \cref{eq:Deltadresidue}].
    \item The remaining spatial integration momenta are scaled by the temperature $T$, which is present as a scale in each term. The projectors in \cref{eq:projectors_L,eq:projectors_T,eq:projectors_C,eq:projectors_D} are independent of the magnitude $K$, so a simple expansion in powers of $K/T$ of the tensor components is possible and produces the $K$-independent term, the first power correction, and so on in a simple manner, following \cite{Gorda:2022fci}.
    \item Following the expansion, the $d$-dimensional $\mathbf{k}$-integrals typically simplify significantly, in particular factorizing into an angular and a radial part, and can be performed with standard methods (see \cite{Gorda:2022fci} and \cref{sec:angints}). 
\end{enumerate}

The exact steps depend on the problem at hand, but we have found the above procedure convenient for the case considered here.

\section{Evaluation of the gluon self-energies}
\label{sec:organization}

In this Section, we provide the details of our general steps above for computing the power corrections $\Pi^{\mu \nu}_{1, \mathrm{Pow}}$ and two-loop HTL corrections $\Pi^{\mu \nu}_\text{2,HTL}$ [see \cref{eq:pi_notation}] for the HTL gluon self-energy. We first show the power corrections and then proceed to the two-loop HTL corrections.

\subsection{One-loop gluon self-energy: the HTL and power correction contributions}

\begin{figure}
\begin{center} 
\includegraphics[scale=0.8]{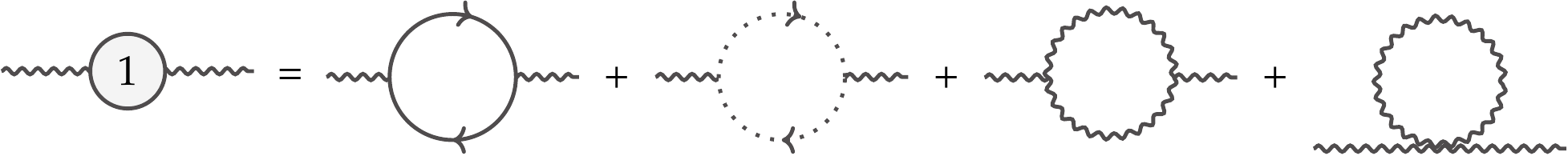} \\[3ex]
\includegraphics[scale=1.1]{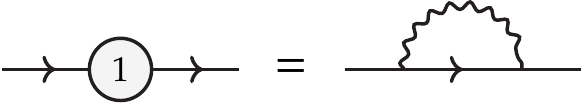}
$\qquad$
\includegraphics[scale=1.1]{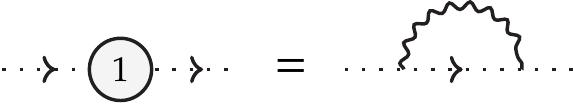}
\end{center}
\caption{%
(Top) The diagrammatic contributions to the one-loop gluon self-energy. (Bottom) diagrammatic contributions to the one-loop fermionic and ghost self-energies. All of these appear as inserts within the two-loop gluon self-energy below.
\label{fig:1loopSE}
}
\end{figure}

At the one-loop level in QCD with general kinematics, there are four diagrams contributing to the self-energy of the gluon. They are shown in the top row of \cref{fig:1loopSE}. The resulting self-energy can be divided into quark and gluonic (or Yang--Mills) contributions
\begin{equation}
    \Pi^{\mathrm {1-loop}}_{\mu\nu}  = \Pi^{\mathrm {1-loop}}_{\mu\nu,\mathrm{qu}} + \Pi^{\mathrm {1-loop}}_{\mu\nu,\mathrm{gl}},
\end{equation}
where the quark contribution arises from the first diagram of \cref{fig:1loopSE}, and the gluonic contributions from the remaining three. The quark contribution is, up to a representation-theoretical factor, identical to that encountered in QED \cite{Manuel:2016wqs,Carignano:2017ovz,Carignano:2019ofj,Gorda:2022fci}. Using the $r/a$-basis and following the same steps as presented in \cite{Gorda:2022fci}, one obtains the following expression for the quark contribution
\begin{equation}
\begin{split}
-i\Pi^{\mathrm {1-loop}}_{\mu\nu,\mathrm{qu}}(K)\delta^{ab} & = -\nf 
    \raisebox{-0.42\height}{\includegraphics[height=1.5cm]{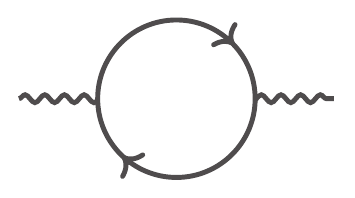}}\\
& = - \gs^2\nf \frac{\delta^{ab}}{2}\int_P \Delta^d(P) F_{\mu\nu}^{\mathrm{qu}} \biggl \{N^{-}_F(P) + N^{+}_F(P)   \biggr \}\Delta^R(K+P),
\end{split}
\end{equation}
where the function $F_{\mu\nu}^{\mathrm{qu}}$ is defined via the relation 
\begin{equation}
\gs^2 \frac{\delta^{ab}}{2}F_{\mu\nu}^{\mathrm{qu}}  \equiv \mathrm{Tr}\left[\left(iV^a_{\mu}\right) \slashed{P} \left(iV^b_{\nu}\right) (\slashed{K}+\slashed{P}) \right],
\label{eq:quarknumerator}
\end{equation}
with the quark-gluon vertices given in \cref{sec:appA}.
The corresponding $\Pi^{\mathrm {1-loop}}_{\mathrm{T},\mathrm{qu}}, \Pi^{\mathrm {1-loop}}_{\mathrm{L},\mathrm{qu}}, \Pi^{\mathrm {1-loop}}_{\mathrm{C},\mathrm{qu}}$ and $\Pi^{\mathrm {1-loop}}_{\mathrm{D},\mathrm{qu}}$ components are then projected out from the full $\mu\nu$-tensor by using \cref{eq:gluonselfcompexpl}. We note that the contributing integrals in quark self-energy components $\mathrm{C}$ and $\mathrm{D}$ vanish due to symmetries.

Following the steps introduced in the previous \cref{subsec:HTLlimit}, we obtain the well-known HTL-limit 
\begin{align}
    \Pi_{\mathrm{T},\mathrm{qu}}^{1,\mathrm{HTL}} & = \frac{\gs^2\nf T^2}{12}\left(1+12\bar\mu^2\right)\Afunc(K)+O(\epsilon), \\
    \Pi_{\mathrm{L},\mathrm{qu}}^{1,\mathrm{HTL}} & = \frac{\gs^2\nf T^2}{6}\left(1+12\bar\mu^2\right)\Bfunc(K)+O(\epsilon),
\end{align}
where we have introduced the following compact notation: 
\begin{align}
\label{eq:AandBfuncandMub}
    \bar \mu & \equiv \frac{\mu}{2\pi T}, \\ 
    \Afunc(K) &\equiv \frac{k_0^2}{k^2}+\left(1-\frac{k_0^2}{k^2}\right)L(K), \\
    \Bfunc(K) &\equiv \left(1-\frac{k_0^2}{k^2}\right)\Bigl (1-L(K)\Bigr ),
\end{align}
with
\begin{equation}
\label{eq:Lfunc}
L(K) \equiv \frac{k^0}{2k}\ln\left (\frac{k^0 + k}{k^0-k} \right ).    
\end{equation}

We can also obtain the UV-divergent power correction \cite{Gorda:2022fci}
\begin{align}
    \Pi_{\mathrm{T},\mathrm{qu}}^{1,\mathrm{Pow}} & = -\frac{\gs^2\nf}{(4\pi)^2}\frac{4K^2}{3} \biggl \{-\frac{1}{2\varepsilon} - \ln \frac{\Lbar}{4\pi T}-L(K)+\frac{\Afunc(K)}{4}  + \frac{\aleph(w)}{2}\biggr \}+O(\epsilon), \\
    \Pi_{\mathrm{L},\mathrm{qu}}^{1,\mathrm{Pow}} & = -\frac{\gs^2\nf}{(4\pi)^2}\frac{4K^2}{3} \biggl \{-\frac{1}{2\varepsilon} - \ln \frac{\Lbar}{4\pi T}-L(K)+\frac{\Bfunc(K)}{2}  + \frac{\aleph(w)}{2}\biggr \}+O(\epsilon),
\end{align}
where the definition of the function $\aleph(w)$ and its argument $w$ are given in \cref{eq:specialfunctions,eq:wdef}, respectively. The divergence is canceled by the wavefunction renormalization counterterm given in \cref{sec:appA}.

A similar calculation for the gluonic diagrams (simplified by the absence of the chemical potential $\mu$) leads to
\begin{equation}
\begin{split}
-i\Pi^{\mathrm {1-loop}}_{\mu\nu,\mathrm{gl}}\delta^{ab} & =  -1 \raisebox{-0.42\height}{\includegraphics[height=1.5cm]{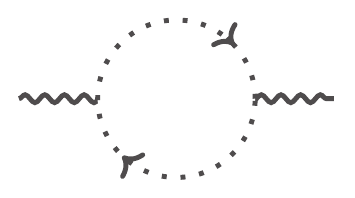}} + \frac{1}{2} \raisebox{-0.42\height}{\includegraphics[height=1.5cm]{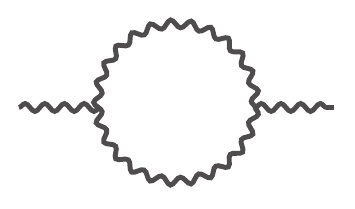}} +\frac{1}{2} \raisebox{-0.42\height}{\includegraphics[height=1.5cm]{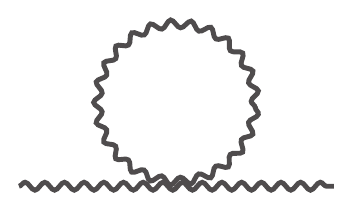}}\\
& = -i\Pi^{\mathrm {1-loop}}_{\mu\nu,\mathrm{gl},0}\delta^{ab} + (-i)\Pi^{\mathrm {1-loop}}_{\mu\nu,\mathrm{gl},\hat\xi}\delta^{ab},
\label{eq:g1loopse1}
\end{split}
\end{equation}
where the gluonic part is further subdivided into parts independent of $\hat \xi \equiv 1-\xi$ (denoted by the subscript 0) and parts proportional to $\hat \xi$ (denoted by the subscript $\hat \xi$), which vanish in Feynman gauge. In the $r/a$-basis, the sum of three gluonic diagrams yields the following expressions,
\begin{equation}
\begin{split}
-i\Pi^{\mathrm {1-loop}}_{\mu\nu,\mathrm{gl},0}\delta^{ab}  = \frac{\gs^2 C_A \delta^{ab}}{2} \int_P \Delta^{rr}_{B,1}(P) \bigg\{2 \left(F_{\mu\nu}^{\mathrm{3g}}-2F_{\mu\nu}^{\mathrm{gh}}\right)  \Delta^R(K+P)  + F_{\mu\nu}^{\mathrm{4g}}  \bigg\},
\label{eq:g0loopse1}
\end{split}
\end{equation}
and
\begin{equation}
\begin{split}
-i\Pi^{\mathrm {1-loop}}_{\mu\nu,\mathrm{gl},\hat\xi}\delta^{ab} & = \frac{-i\hat \xi \gs^2 C_A \delta^{ab}}{2} \int_P \biggl \{\Delta^{rr}_{B,2}(P)F_{\mu\nu}^{\text{3g}_{\xi_1}}\Delta^R(K + P) + \Delta^{rr}_{B,1}(P)F_{\mu\nu}^{\text{3g}_{\xi_2}} [\Delta^R(K + P)]^2  \\ 
&\quad -2i\hat \xi \Delta^{rr}_{B,2}(P)F_{\mu\nu}^{\text{3g}_{\xi_3}} [\Delta^R(K +P)]^2   + \Delta^{rr}_{B,2}(P)F_{\mu\nu}^{\text{4g}_\xi}  \biggr \},
\label{eq:g1hatxiloopse1}
\end{split}
\end{equation}
where the symmetric scalar propagator $\Delta^{rr}_{B,n}$ with $n=1,2$ is defined in \cref{eq:KMSrelation}. 
The functions $F_{\mu\nu}^{\mathrm{3g}}, F_{\mu\nu}^{\mathrm{gh}}$ and $F_{\mu\nu}^{\mathrm{4g}}$ in \cref{eq:g0loopse1} are defined via relations:
\begin{equation}  
\begin{split}
\gs^2 C_A \delta^{ab}F_{\mu\nu}^{\mathrm{3g}} & \equiv \left (iV^{acd}_{\mu\rho\lambda}(-K,-P,K+P)\right )  \left (iV^{bcd}_{\nu\sigma\gamma}(K,P,-K-P)\right ) g^{\rho\sigma} g^{\gamma\lambda}, \\
\gs^2 C_A \delta^{ab}F_{\mu\nu}^{\mathrm{gh}} & \equiv \left (iV^{dac}_{\mu}(K+P)\right ) \left (iV^{cbd}_{\nu}(P)\right ), \\
\gs^2 C_A \delta^{ab}F_{\mu\nu}^{4g} & \equiv \left (iV^{abcc}_{\mu\nu\rho\sigma}\right ) g^{\rho\sigma}. 
\label{eq:gluonnumerators}
\end{split}
\end{equation}
Similarly, the $\hat\xi$-dependent functions $F_{\mu\nu}^{\text{3g}_{\xi_1}}$, $F_{\mu\nu}^{\text{3g}_{\xi_2}}$, $F_{\mu\nu}^{\text{3g}_{\xi_3}}$ and $F_{\mu\nu}^{\text{4g}_\xi}$ in \cref{eq:g1hatxiloopse1} are defined as:
\begingroup
\allowdisplaybreaks
\begin{equation}
\label{eq:gluonnumeratorgaugedep}
\begin{split}
\gs^2 C_A \delta^{ab}F_{\mu\nu}^{\text{3g}_{\xi_1}} & \equiv \left (iV^{acd}_{\mu\rho\lambda}(-K,K+P,-P)\right )  \left (iV^{bcd}_{\nu\sigma\gamma}(K,-K-P,P)\right ) g^{\rho\sigma} P^\gamma P^\lambda\\
& + \left(iV^{acd}_{\mu\rho\lambda}(-K,-P,K+P)\right)  \left (iV^{bcd}_{\nu\sigma\gamma}(K,P,-K-P)\right ) g^{\gamma\lambda} P^\rho P^\sigma, \\
\gs^2 C_A \delta^{ab} F_{\mu\nu}^{\text{3g}_{\xi_2}} &\equiv \left (iV^{acd}_{\mu\rho\lambda}(-K,-P,K+P)\right )  \left (iV^{bcd}_{\nu\sigma\gamma}(K,P,-K-P)\right )  g^{\rho\sigma} (K + P)^\gamma (K + P)^\lambda \\
& + \left (iV^{acd}_{\mu\rho\lambda}(-K,K+P,-P)\right)  \left (iV^{bcd}_{\nu\sigma\gamma}(K,-K-P,P)\right) g^{\gamma\lambda} (K+P)^\rho (K+P)^\sigma, \\
\gs^2 C_A \delta^{ab} F_{\mu\nu}^{\text{3g}_{\xi_3}} &\equiv\left (iV^{acd}_{\mu\rho\lambda}(-K,-P,K+P)\right )  \left (iV^{bcd}_{\nu\sigma\gamma}(K,P,-K-P)\right ) \\
&\quad\times P^\rho P^\sigma (K + P)^\gamma (K + P)^\lambda, \\
\gs^2 C_A \delta^{ab} F_{\mu\nu}^{\text{4g}_\xi} & \equiv \left (iV^{abcc}_{\mu\nu\rho\sigma}\right ) P^{\rho}P^{\sigma}.
\end{split}
\end{equation}
\endgroup

In the end, the gluonic diagrams give for the HTL and first power corrections
\begin{align}
\Pi_{\mathrm{T},\mathrm{gl}}^{1,\mathrm{HTL}} & = \frac{\gs^2\ca T^2}{6}\Afunc(K)+O(\epsilon), \\
\Pi_{\mathrm{L},\mathrm{gl}}^{1,\mathrm{HTL}} & = \frac{\gs^2\ca T^2}{3}\Bfunc(K)+O(\epsilon),
\end{align}
and
\begin{align}
\begin{split}
\Pi_{\mathrm{T},\mathrm{gl}}^{1,\mathrm{Pow}} & = -\frac{\gs^2\ca}{(4\pi)^2}\frac{K^2}{12} \biggl \{4+4\left(10+3\hat\xi\right)\left(\frac{1}{2\varepsilon}+\ln\frac{e^\gamE\Lbar}{4\pi T}\right)+4\left(10-3\hat\xi\right)L(K) \\
&\quad-\left(4-3\hat\xi^2\right)\Afunc(K)\biggr \}+O(\epsilon),
\end{split} \\
\begin{split}
\Pi_{\mathrm{L},\mathrm{gl}}^{1,\mathrm{Pow}} & = -\frac{\gs^2\ca}{(4\pi)^2}\frac{K^2}{12} \biggl \{4+4\left(10+3\hat\xi\right)\left(\frac{1}{2\varepsilon}+\ln\frac{e^\gamE\Lbar}{4\pi T}\right)-6\hat\xi\left(2+\hat\xi\right) \\
&\quad+2\left(20+3\hat\xi^2\right)L(K)-2\left(4-3\hat\xi^2\right)\Bfunc(K)\biggr \}+O(\epsilon),
\end{split} \\
\Pi_{\mathrm{C},\mathrm{gl}}^{1,\mathrm{Pow}} & =  -\frac{\gs^2\ca}{(4\pi)^2}\frac{K^2}{12} \frac{k^0}{k} \left(6\hat\xi\right)\Bigl(1-L(K)\Bigr)+O(\epsilon), \\
\Pi_{\mathrm{D},\mathrm{gl}}^{1,\mathrm{Pow}} & = 0,
\end{align}
respectively. Again, the UV-divergences in the power corrections are canceled by wavefunction renormalization. Notably, as discussed earlier in the Introduction, the component $\propto \mathbb{P}_C$ vanishes only in specific cases, eg. in the HTL limit or in Feynman gauge, but is generally nonzero.

\subsection{Two-loop gluon self-energy: the HTL contribution}\label{sec:twoloopSE}

\begin{figure}
\begin{center}
\includegraphics[scale=0.8]{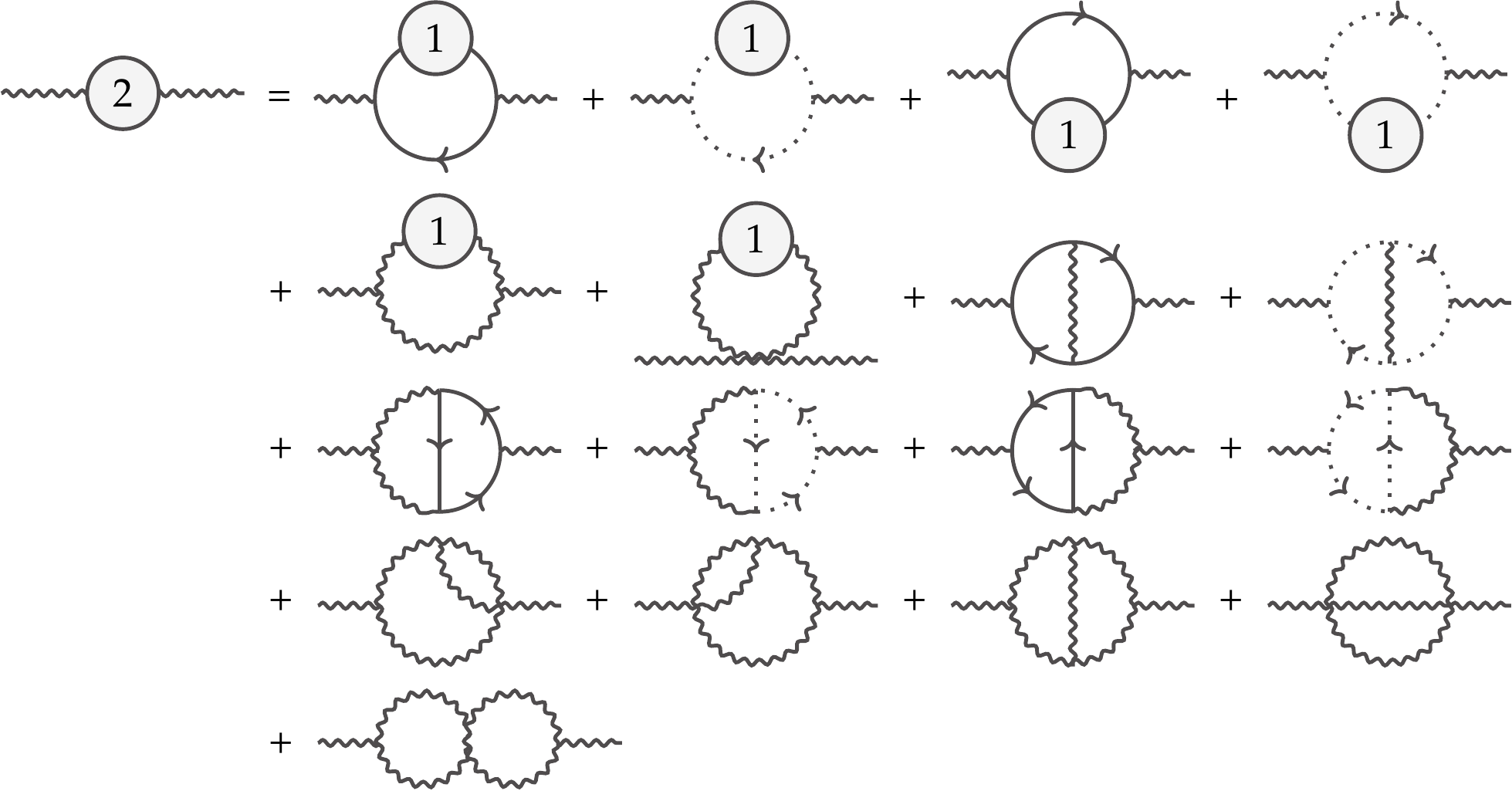}
\end{center}
\caption{The diagrammatic contributions to the two-loop gluon self-energy, with one-loop gluonic and fermionic insertions as in \cref{fig:1loopSE}.}
\label{fig:2loopSE}
\end{figure}

There are 23 two-loop diagrams that contribute to the gluon HTL self-energy at NLO (see e.g.~\cite{Kajantie:2001hv}). We have grouped the diagrams with a one-loop self-energy insertion together in \cref{fig:2loopSE}, with the one-loop self-energy diagrams that appear as inserts displayed in \cref{fig:1loopSE}. Note that some diagrams come with mirrored counterparts, and for those for which they are not equal, we show them explicitly in what follows.

We again split the two-loop part of the NLO HTL gluon self-energy into two terms  
\begin{equation}
\Pi^{\mathrm{2-loop}}_{\mu\nu} = \Pi^{\mathrm{2-loop}}_{\mu\nu, \mathrm{qu}} + \Pi^{\mathrm{2-loop}}_{\mu\nu,\mathrm{gl}},  
\end{equation}
where $\Pi^{\mathrm{2-loop}}_{\mu\nu, \mathrm{qu}}=\sum_i \Pi^{\mathrm{2-loop}}_{\mu\nu, \mathrm{qu}_i}$ contains the two-loop fermionic self-energy diagrams $\mathrm{qu}_i$ (linearly proportional to $\nf$) and $\Pi^{\mathrm{2-loop}}_{\mu\nu, \mathrm{gl}}=\sum_i \Pi^{\mathrm{2-loop}}_{\mu\nu, \mathrm{gl}_i}$ contains all the gluonic two-loop self-energy diagrams $\mathrm{gl}_i$.

In the $r/a$-basis of the real-time formalism, each diagram is associated with a causal labeling (coloring) of the lines, encoded as $\mathcal{C}=\lbrace{c_i\rbrace}_{i=1}^{E}$ where $E$ is the number of (internal) lines in the diagram, and each $c_i$ determines the $r/a$-designation associated with a given line. The admissible causal labelings for each topology of the two-loop self-energy are shown in \cref{sec:ralabels}, and the complete contribution of a standard Feynman diagram as drawn here is obtained by summing over the causal labelings \cite{Ghiglieri:2020dpq}, which we denote by $\sum_{\mathcal{C}}$.

We start by employing the Feynman rules found in \cref{sec:appA} and writing the fermionic contributions as
\begingroup
\allowdisplaybreaks
\begin{align}
\begin{split}
-i\Pi^{\mathrm{2-loop}}_{\mu\nu,\mathrm{qu}_{1}}\delta^{ab} &\equiv -\nf
    \raisebox{-0.42\height}{\includegraphics[height=1.5cm]{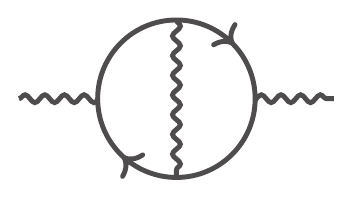}} \\
& = - \nf\sum_{\mathcal{C}}\int_{PQ} D^{\rho\sigma}_{c_{1}}(Q) \mathrm{Tr} \Bigl[ \big(i V_{\mu}^a\big) S_{c_{2}}(P) \big(i V_{\rho}^c\big) S_{c_{3}}(PQ) \\
&\quad\times\big(i V_{\nu}^b\big) S_{c_{4}}(KPQ) \big(i V_{\sigma}^c\big) S_{c_{5}}(KP) \Bigr],
\end{split}\label{eq:qu1diag} \\
\begin{split}
-i\Pi^{\mathrm{2-loop}}_{\mu\nu,\mathrm{qu}_{2a}}\delta^{ab} &\equiv   
    -\nf\scalebox{1}[-1]{\raisebox{-0.59\height}{\includegraphics[height=1.5cm]{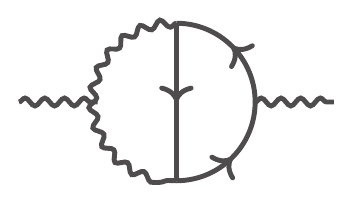}}} \\
& =-\nf \sum_{\mathcal{C}}\int_{PQ} \bigl(iV_{\mu\rho\sigma}^{acd}(-K,-P,KP)\bigr) D^{\rho\lambda}_{c_1}(P) D^{\sigma\gamma}_{c_2}(KP) \\
&\quad\times\mathrm{Tr} \Bigl[ \big(i V_{\lambda}^c\big) S_{c_3}(PQ) \big(i V_{\nu}^b\big) S_{c_4}(KPQ) \big(i V_{\gamma}^d\big) S_{c_5}(Q) \Bigr],
\end{split} \\
\begin{split}
-i\Pi^{\mathrm{2-loop}}_{\mu\nu,\mathrm{qu}_{2b}}\delta^{ab} &\equiv  
    -\nf\scalebox{-1}[1]{\raisebox{-0.42\height}{\includegraphics[height=1.5cm]{figures/SE-2loop/pg6.pdf}}} \\
& = - \nf\sum_{\mathcal{C}}\int_{PQ}\bigl(iV_{\nu\rho\sigma}^{bcd}(K,-KP,P)\bigr) D^{\rho\lambda}_{c_1}(KP) D^{\sigma\gamma}_{c_2}(P) \\
&\quad\times\mathrm{Tr} \Bigl[ \big(i V_{\mu}^a\big) S_{c_3}(PQ) \big(i V_{\gamma}^d\big) S_{c_4}(Q) \big(i V_{\lambda}^c\big) S_{c_5}(KPQ) \Bigr],
\end{split} \\
\begin{split}
-i\Pi^{\mathrm{2-loop}}_{\mu\nu,\mathrm{qu}_{3a}}\delta^{ab} &\equiv  
    -\nf\raisebox{-0.42\height}{\includegraphics[height=1.5cm]{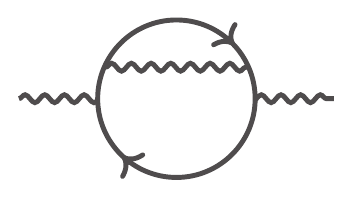}} \\
& = -\nf \sum_{\mathcal{C}}\int_{PQ} D^{\rho\sigma}_{c_1}(Q) \mathrm{Tr} \Bigl[ \big(i V_{\mu}^a\big) S_{c_2}(P) \big(i V_{\nu}^b\big) S_{c_3}(KP) \\
&\quad\times\big(i V_{\rho}^c\big) S_{c_4}(KPQ) \big(i V_{\sigma}^c\big) S_{c_5}(KP) \Bigr],
\end{split} \\
\begin{split}
-i\Pi^{\mathrm{2-loop}}_{\mu\nu,\mathrm{qu}_{3b}}\delta^{ab} &\equiv  
    -\nf\scalebox{-1}[1]{\raisebox{-0.42\height}{\includegraphics[height=1.5cm]{figures/SE-2loop/pg7.pdf}}} \\
& =  - \nf\sum_{\mathcal{C}}\int_{PQ} D^{\rho\sigma}_{c_1}(Q) \mathrm{Tr} \Bigl[ \big(i V_{\mu}^a\big) S_{c_2}(P) \big(i V_{\rho}^c\big) S_{c_3}(PQ) \\
&\quad\times\big(i V_{\sigma}^c\big) S_{c_4}(P) \big(i V_{\nu}^b\big) S_{c_5}(KP) \Bigr],
\end{split} \\
\begin{split}
-i\Pi^{\mathrm{2-loop}}_{\mu\nu,\mathrm{qu}_{4}}\delta^{ab} &\equiv  
    -\nf\raisebox{-0.32\height}{\includegraphics[height=1.5cm]{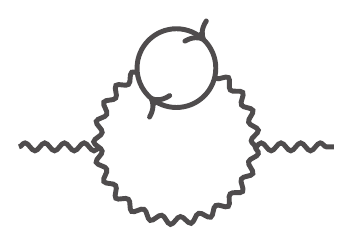}} \\
& = - \nf\sum_{\mathcal{C}}\int_{PQ} \bigl(iV_{\mu\rho\sigma}^{acd}(-K,-P,KP)\bigr)D^{\rho\lambda}_{c_1}(P)\bigl(iV_{\nu\lambda\gamma}^{bce}(K,P,-KP)\bigr) \\
&\quad\times D^{\beta\gamma}_{c_2}(KP) D^{\alpha\sigma}_{c_3}(KP) \mathrm{Tr} \Bigl[\big(i V_{\alpha}^d\big) S_{c_4}(Q) \big(i V_{\beta}^e\big) S_{c_5}(KPQ) \Bigr],
\end{split} \\
\begin{split}
-i\Pi^{\mathrm{2-loop}}_{\mu\nu,\mathrm{qu}_{5}}\delta^{ab} &\equiv 
    -\frac{\nf}{2}\raisebox{-0.42\height}{\includegraphics[height=1.5cm]{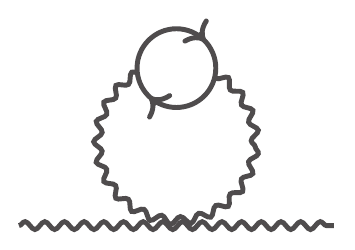}} \\
& = - \frac{\nf}{2}\sum_{\mathcal{C}}\int_{PQ} \bigl(iV_{\mu\nu\rho\sigma}^{abcd}\bigr) D^{\rho\lambda}_{c_1}(Q) D^{\gamma \sigma}_{c_2}(Q) \mathrm{Tr} \Bigl[ \big(i V_{\lambda}^c\big) S_{c_3}(P) \big(i V_{\gamma}^d\big) S_{c_4}(PQ) \Bigr],
\end{split}\label{eq:qu5diag}
\end{align}
\endgroup
where we have abbreviated $KPQ=K+P+Q$ etc., and the trace is taken over both Dirac and fundamental color indices. The color factors associated with these diagrams are given by
\begin{equation}
\mathrm{qu}_1: \ca-2\cf, \qquad
\mathrm{qu}_2: \ca, \qquad
\mathrm{qu}_3: \cf, \qquad
\mathrm{qu}_4: \ca, \qquad
\mathrm{qu}_5: \ca.
\end{equation}

Before proceeding into the evaluation of the integrals in \crefrange{eq:qu1diag}{eq:qu5diag}, one must take care of intermediate pinch singularities \cite{Gorda:2022fci}.
They arise when an integration contour gets squeezed between two poles; this happens when $\eta\to 0$ in terms containing products of retarded and advanced propagators with the same arguments, i.e.~$\Delta^R(P)\Delta^A(P)$. In our case, such terms cancel when applying the relation
\begin{equation}\label{eq:2distrelf}
    N_F^\pm(P_1)N_F^\mp(P_2)
  + N_F^\mp(P_2)N_B^{ }(P_3)
  + N_B^{ }(P_3)N_F^\pm(P_1)+\frac{1}{4}=0,
\end{equation}
where $\sum_i P_i=0$. Once the pinch singularities have been cleared out, we carry on by rewriting the rest of the terms in such a way that each distribution function $N_i(P)$ multiplies a $\Delta^d_n(P)$-propagator with the same argument $P$, ensuring that distributions depend only on radial variables after 0-component integration. Finally, before evaluating the integrals, it is convenient to shift the integration variables so that $\Delta^d_n$-propagators depend only on a single loop momentum.

From this point on, we follow the list of steps outlined in \cref{subsec:HTLlimit} to obtain the HTL limit of the fermionic integrals. In the final step, where we evaluate the angular integrals, it is noteworthy that---just like in QED---all collinear divergences cancel\footnote{In fact, all type-$\Act$ integrals cancel.} after applying the reduction formulas in \cref{sec:angints}. The unrenormalized results for the fermionic part of the self-energy components in the HTL limit then read
\begingroup
\allowdisplaybreaks
\begin{align}
\begin{split}\label{eq:2HTLTbarequ}
\Pi^{2,\mathrm{HTL}}_{\mathrm{T,qu}} &= \frac{g_s^4\nf}{(4\pi)^2}\frac{T^2}{24}\Biggl\{\ca\Bigl(1+12\bar{\mu}^2\Bigr)\biggl[2\left(4-\hat{\xi}\right)\left(\frac{1}{2\epsilon}+1+2\ln\frac{e^{\gamE/2} \Lbar}{4\pi T}\right)\Afunc(K) \\
&\quad-\left(4-\hat{\xi}\right)H(K)+2\left\{4+\left(4+\hat{\xi}\right)\Afunc(K)\right\}L(K)-\hat{\xi}^2A^2(K)\biggr] \\
&\quad+24\ca\left(4-\hat{\xi}\right)\aleph(1,w)\Afunc(K)-24\cf\Bigl(1+4\bar{\mu}^2\Bigr)L(K)\Biggr\} +O(\epsilon),
\end{split} \\
\begin{split}\label{eq:2HTLLbarequ}
\Pi^{2,\mathrm{HTL}}_{\mathrm{L,qu}} &= \frac{g_s^4\nf}{(4\pi)^2}\frac{T^2}{24}\Biggl\{2\ca\Bigl(1+12\bar{\mu}^2\Bigr)\biggl[4+2\left(4-\hat{\xi}\right)\left(\frac{1}{2\epsilon}+\frac{1}{2}+2\ln\frac{e^{\gamE/2} \Lbar}{4\pi T}\right)\Bfunc(K) \\
&\quad+\left(4-\hat{\xi}\right)H(K)+2\hat\xi\left(1+\hat\xi\right)\Bfunc(K) \\
&\quad+2\left(4-\hat\xi^2\right)\Bfunc(K)L(K)-2\hat{\xi}^2B^2(K)\biggr] \\
&\quad+48\ca\left(4-\hat{\xi}\right)\aleph(1,w)\Bfunc(K)+96\Bigl(\ca-2\cf\Bigr)\bar\mu^2\Bfunc(K)\Bigl(1-L(K)\Bigr) \\
&\quad-24\cf\Bigl(1+4\bar{\mu}^2\Bigr)\Biggr\}+O(\epsilon),
\end{split} \\
\begin{split}\label{eq:2HTLCbarequ}
\Pi^{2,\mathrm{HTL}}_{\mathrm{C,qu}} &= \frac{g_s^4\nf}{(4\pi)^2}\frac{T^2}{24}\frac{k^0}{k}\ca\Bigl(1+12\bar{\mu}^2\Bigr)\Bigl(-2\hat{\xi}\Bfunc(K)\Bigr)\Bigl(1-L(K)\Bigr)+O(\epsilon),
\end{split} \\
\Pi^{2,\mathrm{HTL}}_{\mathrm{D,qu}} &= 0,\label{eq:2HTLDbarequ}
\end{align}
\endgroup
where
\begin{equation}
    H(K) \equiv \frac{K^2}{k^2}\Biggl\{\left(2+\ln\frac{K^2}{4k^2}\right)L(K)-\frac{k^0}{2k}\left(\mathrm{Li}_2\frac{k^0+k}{k^0-k}-\mathrm{Li}_2\frac{k^0-k}{k^0+k}\right)\Biggr\}.
\end{equation}
The $1/\epsilon$ divergences in the T and L components above are removed by renormalization, and the appropriate counterterm contributions are obtained by using the Feynman rules in \cref{sec:appA},
\begin{align}
\begin{split}
\Pi^{2,\mathrm{HTL}}_{\mathrm{T,qu,ct}} &= -\frac{g_s^4\nf}{(4\pi)^2}\frac{T^2}{24}\Biggl\{\ca\Bigl(1+12\bar{\mu}^2\Bigr)\biggl[2\left(4-\hat{\xi}\right)\left(\frac{1}{2\epsilon}+1+\ln\frac{\Lbar}{4\pi T}\right)\Afunc(K) \\
&\quad -\left(4-\hat{\xi}\right)H(K)\biggr] +24\ca\left(4-\hat{\xi}\right)\aleph(1,w)\Afunc(K)\Biggr\}+O(\epsilon),
\end{split} \\
\begin{split}
\Pi^{2,\mathrm{HTL}}_{\mathrm{L,qu,ct}} &= -\frac{g_s^4\nf}{(4\pi)^2}\frac{T^2}{24}\Biggl\{2\ca\Bigl(1+12\bar{\mu}^2\Bigr)\biggl[2\left(4-\hat{\xi}\right)\left(\frac{1}{2\epsilon}+\frac{1}{2}+\ln\frac{\Lbar}{4\pi T}\right)\Bfunc(K) \\
&\quad+\left(4-\hat{\xi}\right)H(K)\biggr]+48\ca\left(4-\hat{\xi}\right)\aleph(1,w)\Bfunc(K)\Biggr\}+O(\epsilon),
\end{split}
\end{align}
which clearly cancel the UV-divergences in \crefrange{eq:2HTLTbarequ}{eq:2HTLDbarequ}.

Next, we continue to the evaluation of the gluon contributions $\Pi^{\mathrm{2-loop}}_{\mu\nu,\mathrm{gl}}$ in the HTL limit. To this end, we follow the same procedure as in the fermionic case above, starting with the 16 gluonic two-loop contributions listed in \cref{sec:2loopgluonparts}. They all share the same color factor, namely $\ca^2$. As in the fermionic counterpart, the starting expressions contain pinch singularities, which in this case can be shown to cancel by making use of the purely bosonic relation [cf.~\cref{eq:2distrelf}]
\begin{equation}\label{eq:2distrelb}
    N_B(P_1)N_B(P_2)
  + N_B(P_2)N_B(P_3)
  + N_B(P_3)N_B(P_1)+\frac{1}{4}=0,
\end{equation}
where $\sum_i P_i=0$. After the pinch singularities have been sorted out, the rest of the computation follows the recipe established in the fermionic case, leading to the unrenormalized results
\begingroup
\allowdisplaybreaks
\begin{align}
\begin{split}
\Pi^{2,\mathrm{HTL}}_{\mathrm{T,gl}} &= \frac{g_s^4\ca^2}{(4\pi)^2}\frac{T^2}{12}\Biggl\{2\left(4-\hat{\xi}\right)\left(\frac{1}{2\epsilon}+\frac{1}{2}+\frac{\zeta'(-1)}{\zeta(-1)}+2\ln\frac{e^{\gamE/2} \Lbar}{4\pi T}\right)\Afunc(K) \\
&\quad-\left(4-\hat{\xi}\right)H(K)+2\left\{4+\left(4+\hat{\xi}\right)\Afunc(K)\right\}L(K)-\hat{\xi}^2A^2(K)\Biggr\} +O(\epsilon),
\end{split} \\
\begin{split}
\Pi^{2,\mathrm{HTL}}_{\mathrm{L,gl}} &= \frac{g_s^4\ca^2}{(4\pi)^2}\frac{T^2}{12}\Biggl\{8+4\left(4-\hat{\xi}\right)\left(\frac{1}{2\epsilon}+\frac{\zeta'(-1)}{\zeta(-1)}+2\ln\frac{e^{\gamE/2} \Lbar}{4\pi T}\right)\Bfunc(K) \\
&\quad+2\left(4-\hat{\xi}\right)H(K)+4\hat\xi\left(1+\hat\xi\right)\Bfunc(K) \\
&\quad+4\left(4-\hat\xi^2\right)\Bfunc(K)L(K)-4\hat{\xi}^2B^2(K)\Biggr\}+O(\epsilon),
\end{split} \\
\begin{split}
\Pi^{2,\mathrm{HTL}}_{\mathrm{C,gl}} &=  \frac{g_s^4\ca^2}{(4\pi)^2}\frac{T^2}{12}\frac{k^0}{k}\Bigl(-2\hat{\xi}\Bfunc(K)\Bigr)\Bigl(1-L(K)\Bigr)+O(\epsilon),
\end{split}\\
\Pi^{2,\mathrm{HTL}}_{\mathrm{D,gl}} &= 0.
\end{align}
\endgroup
The UV-divergences in the T and L components are again canceled by the counterterm contributions which read
\begingroup
\allowdisplaybreaks
\begin{align}
\begin{split}
\Pi^{2,\mathrm{HTL}}_{\mathrm{T,gl,ct}} &= -\frac{g_s^4\ca^2}{(4\pi)^2}\frac{T^2}{12}\Biggl\{2\left(4-\hat{\xi}\right)\left(\frac{1}{2\epsilon}+\frac{1}{2}+\frac{\zeta'(-1)}{\zeta(-1)}+\ln\frac{\Lbar}{4\pi T}\right)\Afunc(K) \\
&\quad -\left(4-\hat{\xi}\right)H(K)\Biggr\} +O(\epsilon),
\end{split} \\
\begin{split}
\Pi^{2,\mathrm{HTL}}_{\mathrm{L,gl,ct}} &= -\frac{g_s^4\ca^2}{(4\pi)^2}\frac{T^2}{12}\Biggl\{4\left(4-\hat{\xi}\right)\left(\frac{1}{2\epsilon}+\frac{\zeta'(-1)}{\zeta(-1)}+\ln\frac{\Lbar}{4\pi T}\right)\Bfunc(K) \\
&\quad+2\left(4-\hat{\xi}\right)H(K)\Biggr\}+O(\epsilon).
\end{split}
\end{align}
\endgroup

\section{Results and discussion}
\label{sec:summaryoftheres}

Following the computations discussed in the previous Sections, we are left with the renormalized NLO HTL gluon self-energies at finite temperature and chemical potential and in an arbitrary $R_\xi$ gauge\footnote{In all cases, the $D$-component of the self-energy, proportional to $K^\mu K^\nu$, vanishes, and we do not write it explicitly.}: First, the contributions from two-loop diagrams with a strict HTL limit:
\begingroup
\allowdisplaybreaks
\begin{align}
\begin{split}
\Pi^{2,\mathrm{HTL}}_\mathrm{T} &= \frac{g_s^4(\Lbar)}{(4\pi)^2}\frac{T^2}{24}\Biggl\{\left[2\ca^2+\ca\nf\Bigl(1+12\bar{\mu}^2\Bigr)\right]\biggl[2\left(4-\hat{\xi}\right)\Afunc(K)\ln\frac{e^\gamE \Lbar}{4\pi T} \\
&\quad+2\left\{4+\left(4+\hat{\xi}\right)\Afunc(K)\right\}L(K)-\hat{\xi}^2A^2(K)\biggr] \\
&\quad-24\cf\nf\Bigl(1+4\bar{\mu}^2\Bigr)L(K)\Biggr\}+O(\epsilon),
\end{split} \\
\begin{split}
\Pi^{2,\mathrm{HTL}}_\mathrm{L} &= \frac{g_s^4(\Lbar)}{(4\pi)^2}\frac{T^2}{24}\Biggl\{\left[2\ca^2+\ca\nf\Bigl(1+12\bar{\mu}^2\Bigr)\right]\biggl[8+4\left(4-\hat{\xi}\right)\Bfunc(K)\ln\frac{e^\gamE \Lbar}{4\pi T} \\
&\quad+4\hat\xi\left(1+\hat\xi\right)\Bfunc(K)+4\left(4-\hat{\xi}^2\right)\Bfunc(K)L(K)-4\hat\xi^2B^2(K)\biggr] \\
&\quad+96\Bigl(\ca-2\cf\Bigr)\nf\bar{\mu}^2\Bfunc(K)\Bigl(1-L(K)\Bigr)-24\cf\nf\Bigl(1+4\bar{\mu}^2\Bigr)\Biggr\} + O(\epsilon),
\end{split} \\
\begin{split}
\Pi^{2,\mathrm{HTL}}_\mathrm{C} &= \frac{g_s^4(\Lbar)}{(4\pi)^2}\frac{T^2}{24}\frac{k^0}{k}\left[2\ca^2+\ca\nf\Bigl(1+12\bar{\mu}^2\Bigr)\right]\Bigl(-2\hat{\xi}\Bfunc(K)\Bigr)\Bigl(1-L(K)\Bigr) + O(\epsilon),
\end{split}
\end{align}
\endgroup
where $\hat\xi \equiv 1- \xi$ and the functions $A$, $B$ and $L$ are given in \crefrange{eq:AandBfuncandMub}{eq:Lfunc}. Next, the contributions from one-loop diagrams where the first power corrections are considered:
\begingroup
\allowdisplaybreaks
\begin{align}
\begin{split}
\Pi^{1,\mathrm{Pow}}_\mathrm{T} &= -\frac{g_s^2(\Lbar)}{(4\pi)^2}\frac{K^2}{12}\Biggl\{\left[4\ca\Bigl(10+3\hat{\xi}\Bigr)-16\nf\right]\ln\frac{e^\gamE \Lbar}{4\pi T} \\
&\quad+\ca\left[4+4\Bigl(10-3\hat{\xi}\Bigr)L(K)-\Bigl(4-3\hat{\xi}^2\Bigr)\Afunc(K)\right] \\
&\quad+4\nf\Bigl[4\gamE-4L(K)+\Afunc(K)+2\aleph(w)\Bigr] \Biggr\}+O(\epsilon),
\end{split} \\
\begin{split}
\Pi^{1,\mathrm{Pow}}_\mathrm{L} &= -\frac{g_s^2(\Lbar)}{(4\pi)^2}\frac{K^2}{12}\Biggl\{\left[4\ca\Bigl(10+3\hat{\xi}\Bigr)-16\nf\right]\ln\frac{e^\gamE \Lbar}{4\pi T} \\
&\quad+2\ca\left[2-6\hat{\xi}-3\hat{\xi}^2+\Bigl(20+3\hat{\xi}^2\Bigr)L(K)-\Bigl(4-3\hat{\xi}^2\Bigr)\Bfunc(K)\right] \\
&\quad+8\nf\Bigl[2\gamE-2L(K)+\Bfunc(K)+\aleph(w)\Bigr] \Biggr\}+O(\epsilon),
\end{split} \\
\begin{split}
\Pi^{1,\mathrm{Pow}}_\mathrm{C} &= -\frac{g_s^2(\Lbar)}{(4\pi)^2}\frac{K^2}{12}\frac{k^0}{k}\Bigl(6\ca\hat{\xi}\Bigl)\Bigl(1-L(K)\Bigl)+O(\epsilon).
\end{split}
\end{align}
\endgroup

Let us begin with a few cross-checks of the above results. First, our results reproduce the know QED ones in the limit $\ca \to 0$ \cite{Manuel:2016wqs,Carignano:2017ovz,Carignano:2019ofj,Gorda:2022fci}. Furthermore, they match with the recent results obtained in a kinetic-theory approach in Feynman gauge 
\cite{Ekstedt:2023anj,Ekstedt:2023oqb}, although a direct comparison requires a field redefinition \cite{Kajantie:1995dw}. As a further cross-check, we also see agreement at specific known kinematic limits, in particular with the matching coefficient $\alpha_{\mathrm{E6}}$
of the dimensionally reduced theory of Electrostatic QCD \cite{Vuorinen:2003fs,Laine:2005ai} and the associated generalization of the physical Debye screening mass. The specific kinematic limit used here corresponds to setting $k^0=0$ and then taking the limit $k \rightarrow 0$, and the gauge-dependence of the individual terms cancel as the quantity considered is a combination of the power correction and the two-loop correction (specifically, $\Pi^{2\text{,HTL}}-\Pi^{1\text{,HTL}}\Pi^{1,\text{Pow}}$, which is a natural combination appearing after turning the kinetic term canonical \cite{Ekstedt:2023oqb}).

We turn now to a discussion of the results. We find that the computed pieces of the NLO HTL gluon self-energy at $O(\gs^4)$ contain  T, L, and C tensor components, with a vanishing D component. Each of these components is finite after UV-renormalization, which indicates that the sum of the corresponding (soft) resummed $O(g_s^4)$ one- and two-loop contributions must be UV-finite, as any remaining divergence would cancel with IR-divergences of the NLO HTL contributions computed here.

Beyond that, we observe that unlike the well-known leading-order HTL results, the individual components are gauge-dependent, with the C component directly vanishing in Feynman gauge. Here, we remind the reader that there is a missing soft component that must be computed to obtain the complete NNLO soft gluon self-energy (or NLO at small $T$), and it is not clear whether this complete result will be gauge-dependent or not. Naturally, all physical quantities derived from it must be gauge-invariant, but it is not obvious whether the gauge-independence is manifest on the level of the complete (soft) self-energy itself, or in the various combinations and limits in which it appears.

Finally, we close with a small outlook. Computing the remaining soft component of the gluon self-energy is clearly of interest, as it will enable calculating, e.g.~higher-order corrections to the plasmon frequency and other physical quantities. One way of tackling this missing piece would be to generalize our automated code to include the real-time HTL Feynman rules derived in \cite{Caron-Huot:2007cma}. Another interesting direction would be to generalize our present results to zero temperature to compute the mixed pressure of cold quark matter at N3LO, following \cite{Gorda:2021kme}. Such work is ongoing \cite{Gorda:ongoing}.

\section{Acknowledgements}

The authors would like to thank Andreas Ekstedt and Aleksi Vuorinen for enlightening discussions. We are also thankful to Andreas Ekstedt for sending some of his unpublished results for comparison. Lastly, we wish to thank the anonymous referee for their insightful comments which led to an improved discussion on gauge-invariance and the missing soft resummed contributions. This work is supported in part by the Deutsche Forschungsgemeinschaft (DFG, German Research Foundation)--Project ID 279384907--SFB 1245 and by the State of Hesse within the Research Cluster ELEMENTS (Project ID 500/10.006) (T.G.). R.P. and K.S. have been supported by the Academy of Finland grant no.~347499 and 353772 as well as by the European Research Council, grant
no.~725369. S.S. acknowledges support of the DFG cluster of excellence ORIGINS funded by the DFG under Germany’s Excellence Strategy - EXC-2094-390783311. In addition, K.S. gratefully acknowledges support from the Finnish Cultural Foundation.  

\appendix

\section{\texorpdfstring{QCD Feynman rules in covariant $R_\xi$-gauges}{QCD Feynman rules in covariant R\_ξ-gauges}}
\label{sec:appA}

The renormalized gauge-fixed QCD Lagrangian with massless quarks reads
\begin{equation}
\begin{split}
\mathcal{L}_\mathrm{QCD} &= \frac{1}{2}Z_3A^{\mu a}\left(\partial^2 g_{\mu\nu}-\partial_\mu\partial_\nu\right)A^{\nu a} +\frac{1}{2\xi}A^{\mu a}\partial_\mu\partial_\nu A^{\nu a} + Z_{3c}\bar{c}^a\partial^2 c^a + iZ_2\bar{\psi}_i\slashed{\partial}\psi_i \\
&\quad -g_sZ_{A^3}f^{abc}(\partial_\mu A_\nu^a)A^{\mu b}A^{\nu c} - \frac{g_s^2}{4}Z_{A^4}f^{abc}f^{ade}A_\mu^aA_\nu^bA^{\mu d}A^{\nu e} \\
&\quad -g_sZ_{1c}f^{abc}(\partial_\mu \bar{c}^a)A^{\mu b}c^c + g_sZ_1A_\mu^a\bar{\psi}_i\gamma^\mu T_{ij}^a\psi_j , \label{eq:QCDlagrangian}
\end{split}
\end{equation}
where $\xi$ is the gauge-fixing parameter in the $R_\xi$-class of (covariant) gauges, and the counterterms are defined as $\delta_i\equiv Z_i-1$. Next, we write the momentum space Feynman rules corresponding to \cref{eq:QCDlagrangian}. Given the scalar function $\Delta(P)=-i/P^2$ (suppressing the pole prescription $i\eta$ in this Appendix, see \cref{subsec:realtimeQCD} for the $r/a$ basis rules), the propagators for the gluon, ghost, and quark read
\begin{align}
    D_{\mu\nu}^{ab}(P) &= g_{\mu\nu}\delta^{ab}\Delta(P)-i(1-\xi)P_\mu P_\nu \delta^{ab}\Delta(P)^2, \\
    \tilde{D}^{ab}(P) &= \delta^{ab} \Delta(P), \\
    S_{ij}(P) &= -\slashed{P}\delta_{ij}\Delta(P),
\end{align}
respectively. The 3-gluon and 4-gluon interaction vertices are given by
\begin{equation}
\begin{split}
i V_{\mu\nu\rho}^{abc}(P,Q,R) &= g_s f^{abc} \big[(Q-R)_\mu g_{\nu \rho} + (R-P)_\nu g_{\rho\mu} + (P-Q)_\rho g_{\mu\nu} \big] \\
&= \, \raisebox{-0.46\height}{\includegraphics[scale=1.1]{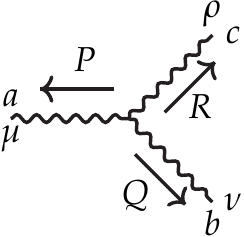}} ,
\end{split}
\end{equation}
and
\begin{equation}
\begin{split}
i V_{\mu\nu\rho\sigma}^{abcd} &= -i g_s^2 \big[f^{abe}f^{cde}(g_{\mu\rho}g_{\nu\sigma}-g_{\mu\sigma}g_{\nu\rho}) \\
& \hspace{1cm}+ f^{ace}f^{dbe}(g_{\mu\sigma}g_{\rho\nu}-g_{\mu\nu}g_{\rho\sigma}) \\
& \hspace{1cm}+ f^{ade}f^{bce}(g_{\mu\nu}g_{\sigma\rho}-g_{\mu\rho}g_{\sigma\nu}) \big] \\
&= \, \raisebox{-0.46\height}{\includegraphics[scale=1.1]{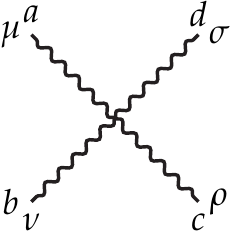}} .
\end{split}
\end{equation}
The ghost--gluon vertex is
\begin{equation}
i V_{\mu}^{abc}(P) = -g_s f^{abc} P_\mu = \, \raisebox{-0.46\height}{\includegraphics[scale=1.1]{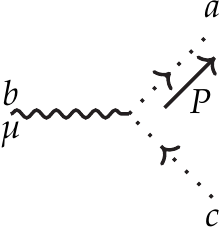}},
\end{equation}
while the quark--gluon vertex reads
\begin{equation}
i V^{a}_{\mu,ij} = i g_s \gamma_\mu T^a_{ij} = \, \raisebox{-0.46\height}{\includegraphics[scale=1.1]{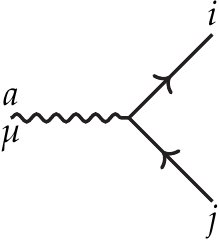}} ,
\end{equation}
using the convention that particle flow is aligned with momentum. The rules for the propagator counterterms are given by
\begin{align}
    i C_{\mu\nu}^{ab}(P) &= -i\delta_3\delta^{ab}(P^2g_{\mu\nu}-P_\mu P_\nu), \\
    i C^{ab}(P) &= -i\delta_{3c}\delta^{ab}P^2, \\
    i C_{ij}(P) &= -i\delta_2\delta_{ij}\slashed{P},
\end{align}
and respectively for the vertices read
\begin{align}
    i C_{\mu\nu\rho}^{abc}(P,Q,R) &= g_s \delta_{A^3} f^{abc} \big[(Q-R)_\mu g_{\nu \rho} + (R-P)_\nu g_{\rho\mu} + (P-Q)_\rho g_{\mu\nu} \big], \\
    \begin{split}
    i C_{\mu\nu\rho\sigma}^{abcd} &= -i g_s^2\delta_{A^4} \big[f^{abe}f^{cde}(g_{\mu\rho}g_{\nu\sigma}-g_{\mu\sigma}g_{\nu\rho}) \\
    &\quad+ f^{ace}f^{dbe}(g_{\mu\sigma}g_{\rho\nu}-g_{\mu\nu}g_{\rho\sigma}) \\
    &\quad+ f^{ade}f^{bce}(g_{\mu\nu}g_{\sigma\rho}-g_{\mu\rho}g_{\sigma\nu}) \big],
    \end{split}\\
    i C_{\mu}^{abc}(P) &= -g_s\delta_{1c} f^{abc} P_\mu, \\
    i C^{a}_{\mu,ij} &= i g_s\delta_1 \gamma_\mu T^a_{ij}.
\end{align}
In the \MSbar\ renormalization scheme, the values of the 1-loop counterterms are found to be \cite{Schwartz:2014sze}:
\begingroup
\allowdisplaybreaks
\begin{align}
    \delta_3 &= \frac{1}{2\epsilon}\frac{g_s^2}{(4\pi)^2}\left[\frac{10}{3}\ca-\frac{4}{3}\nf+(1-\xi)\ca\right], \\
    \delta_{3c} &= \frac{1}{2\epsilon}\frac{g_s^2}{(4\pi)^2}\left[\ca+\frac{1}{2}(1-\xi)\ca\right], \\
    \delta_2 &= \frac{1}{2\epsilon}\frac{g_s^2}{(4\pi)^2}\left[-2\cf+2(1-\xi)\cf\right], \\
    \delta_{A^3} &= \frac{1}{2\epsilon}\frac{g_s^2}{(4\pi)^2}\left[\frac{4}{3}\ca-\frac{4}{3}\nf+\frac{3}{2}(1-\xi)\ca\right], \\
    \delta_{A^4} &= \frac{1}{2\epsilon}\frac{g_s^2}{(4\pi)^2}\left[-\frac{2}{3}\ca-\frac{4}{3}\nf+2(1-\xi)\ca\right], \\
    \delta_{1c} &= \frac{1}{2\epsilon}\frac{g_s^2}{(4\pi)^2}\left[-\ca+(1-\xi)\ca\right], \\
    \delta_1 &= \frac{1}{2\epsilon}\frac{g_s^2}{(4\pi)^2}\left[-2\cf-2\ca+2(1-\xi)\cf+\frac{1}{2}(1-\xi)\ca\right].
\end{align}
\endgroup

\section{Reduction of angular integrals}\label{sec:angints}

In the HTL limit, two-loop integrals factorize into angular and radial parts. Let us recall that the $d$-dimensional spatial integration measure can be correspondingly split as
\begin{equation}
\int\frac{\ud^d\pt}{(2\pi)^d} = \frac{1}{(2\pi)^d}\int \ud \Omega_d (\vt_p) \int_{0}^{\infty} \ud p\, p^{d-1} \equiv \int_{\vt_p} \int_{0}^{\infty} \ud p\, p^{d-1},
\end{equation}
where $\vt_p\equiv \pt/p$ and $\int_{\vt_p}$ denotes angular integration over the $(d-1)$-sphere. With an on-shell loop momentum $V_p\equiv(1,\vt_p)$ and external momentum $K$, the most general angular integral that we encounter in our two-loop (HTL) self-energy calculation reads
\begin{equation}\label{eq:Aabc}
    \Ac_{abc} \equiv \int_{\vt_p\vt_q}(K\cdot V_p)^a (K\cdot V_q)^b (V_p\cdot V_q)^c.
\end{equation}
In the cases where at least one of the indices $\{a,b,c\}$ is a non-negative integer, we may exploit the $d$-dimensional rotational symmetry to reduce \cref{eq:Aabc} into a linear combination of factorized angular integrals. Let us first consider the case $c=0$. Then $\vt_p$ and $\vt_q$ decouple trivially in \cref{eq:Aabc}, leading to a product of the integrals of the type
\begin{equation}
\begin{split}
    \Ac_a &\equiv \int_{\vt_p}(K\cdot V_p)^a = C_d\int_{-1}^1\ud z (1-z^2)^\frac{d-3}{2} (-k^0+k z)^a \\
    &= C_d(-k^0)^a\frac{\Gamma\left(\frac{1}{2}\right)\Gamma\left[\frac{1}{2}(d-1)\right]}{\Gamma\left(\frac{d}{2}\right)} {}_2F_1\left(\frac{1-a}{2},-\frac{a}{2};\frac{d}{2};\frac{k^2}{k_0^2}\right),
\end{split}\label{eq:Aint}
\end{equation}
where the factors coming from integrating over the trivial angles have been absorbed into
\begin{equation}
    C_d \equiv \frac{4}{(4\pi)^\frac{d+1}{2}\Gamma\left(\frac{d-1}{2}\right)}.
\end{equation}
Also, the cases $a=0$ and $b=0$ factorize in a similar fashion, leading us to evaluate the integral
\begin{equation}
\begin{split}
    \Act_a &\equiv \int_{\vt_p}(V_q\cdot V_p)^a = C_d\int_{-1}^1\ud z (1-z^2)^\frac{d-3}{2} (-1+z)^a \\
    &= C_d(-1)^{a+1}2^{d+a-2} \frac{\pi\sec\left[\frac{\pi}{2}(d+2a)\right]\Gamma\left[\frac{1}{2}(d-1)\right]}{\Gamma\left[\frac{1}{2}(3-d-2a)\right]\Gamma(d+a-1)}.
\end{split}\label{eq:Atint}
\end{equation}
Next, consider the case where one of the indices in \cref{eq:Aabc} is a positive integer. Now we can employ rotational symmetry to reduce tensor subintegrals into linear combinations of the scalar integrals in \cref{eq:Aint,eq:Atint}. For instance, if $c=1$ we may use
\begin{equation}
\int_{\vt_p} \vt_p(K\cdot V_p)^a = \vt_k \int_{\vt_p} (\vt_k\cdot\vt_p)(K\cdot V_p)^a = \frac{\vt_k}{k} \left(k^0 \Ac_a+\Ac_{a+1}\right).
\end{equation}
Applying relations such as the one above, we obtain
\begingroup
\allowdisplaybreaks
\begin{align}
    \Ac_{0bc} &= \Ac_b\Act_c, \\
    \Ac_{ab0} &= \Ac_a\Ac_b, \\
    \Ac_{1bc} &= k^0\Ac_b\Act_{c+1}+\Ac_{b+1}\Act_c+\Ac_{b+1}\Act_{c+1}, \\
    \Ac_{ab1} &= \frac{1}{k^2}\left(-K^2\Ac_a\Ac_b+k^0\Ac_a\Ac_{b+1}+k^0\Ac_{a+1}\Ac_b+\Ac_{a+1}\Ac_{b+1}\right), \\
    \begin{split}
    \Ac_{2bc} &= \frac{1}{d-1}\Bigl(-2K^2\Ac_b\Act_{c+1}+(d k_0^2-k^2)\Ac_b\Act_{c+2}+2(d+1)k^0\Ac_{b+1}\Act_{c+1} \\
    &\quad+2d k^0\Ac_{b+1}\Act_{c+2}+(d-1)\Ac_{b+2}\Act_c+2d\Ac_{b+2}\Act_{c+1}+d\Ac_{b+2}\Act_{c+2} \Bigr),
    \end{split} \\
    \begin{split}
    \Ac_{ab2} &= \frac{1}{d-1}\frac{1}{k^4}\Bigl(d K^4\Ac_a\Ac_b-2d k^0K^2\Ac_a\Ac_{b+1}+(d k_0^2-k^2)\Ac_a\Ac_{b+2} \\
    &\quad-2d k^0K^2\Ac_{a+1}\Ac_b+2(2dk_0^2-(d-1)k^2)\Ac_{a+1}\Ac_{b+1}+2d k^0\Ac_{a+1}\Ac_{b+2}  \\
    &\quad+(d k_0^2-k^2)\Ac_{a+2}\Ac_b+2d k^0\Ac_{a+2}\Ac_{b+1}+d\Ac_{a+2}\Ac_{b+2}\Bigr).
    \end{split}
\end{align}
\endgroup
It turns out that every angular integral we need in our two-loop calculation can be reduced using the above formulas.

\section{\texorpdfstring{$r/a$ assignments}{r/a assignments}}\label{sec:ralabels}
Below, we list the $r/a$ labelings for each topology required in the computation. Note that the nature of any given line (i.e.~gluonic or fermionic) is not relevant for these labelings.

\begin{center}
    \includegraphics[scale=0.8]{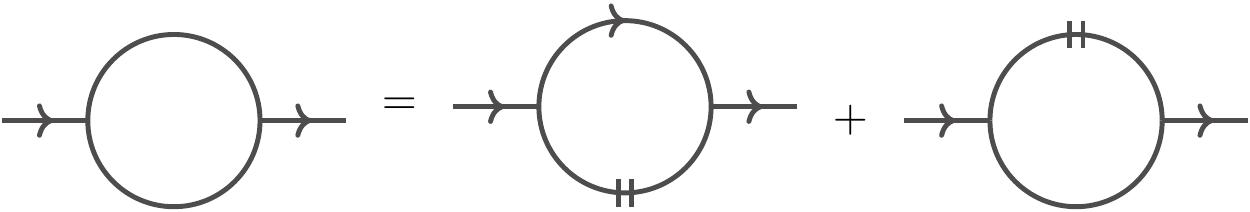}
\end{center}

\begin{center}
    \includegraphics[scale=0.8]{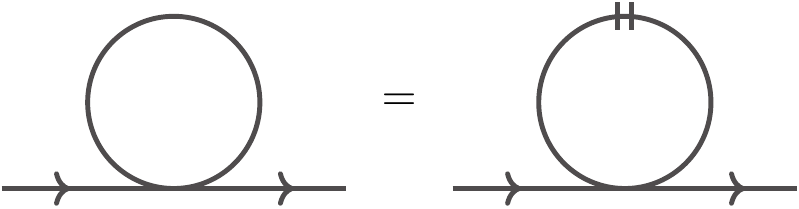}
\end{center}

\begin{center}
    \includegraphics[scale=0.8]{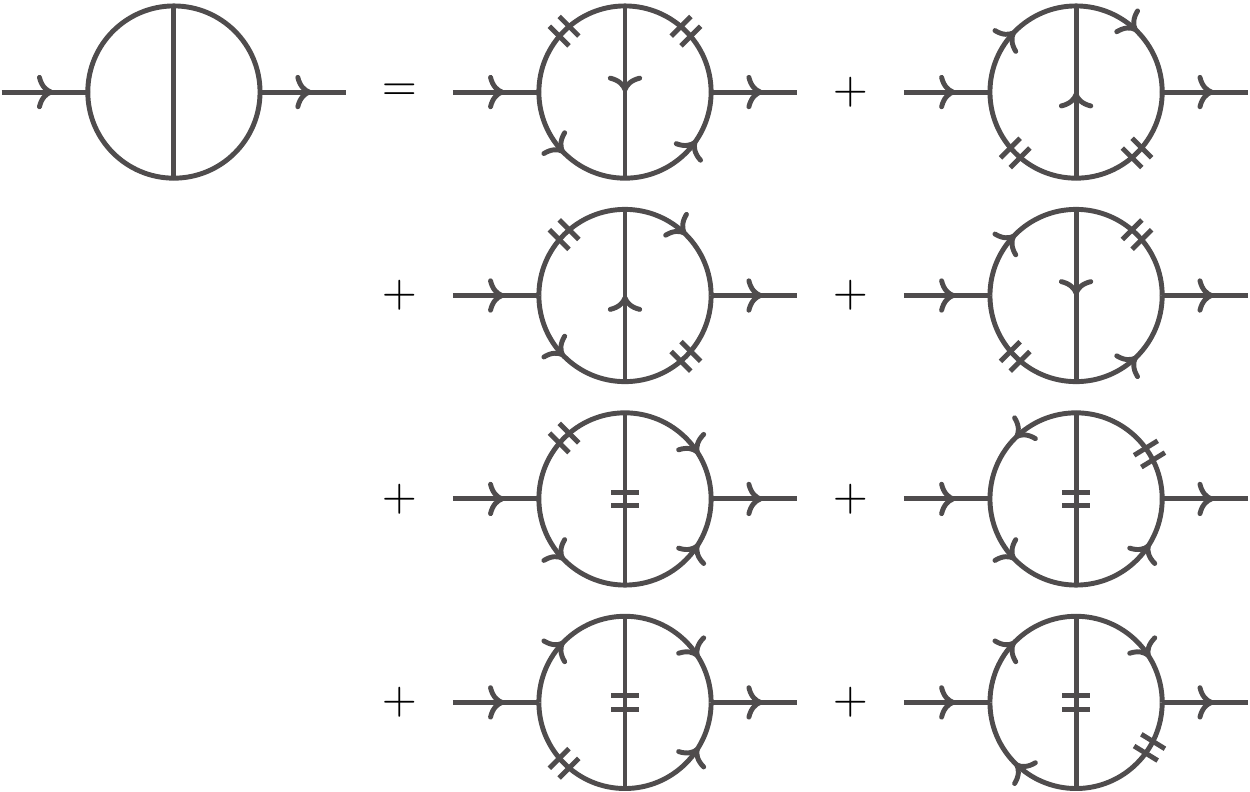}
\end{center}

\begin{center}
    \includegraphics[scale=0.8]{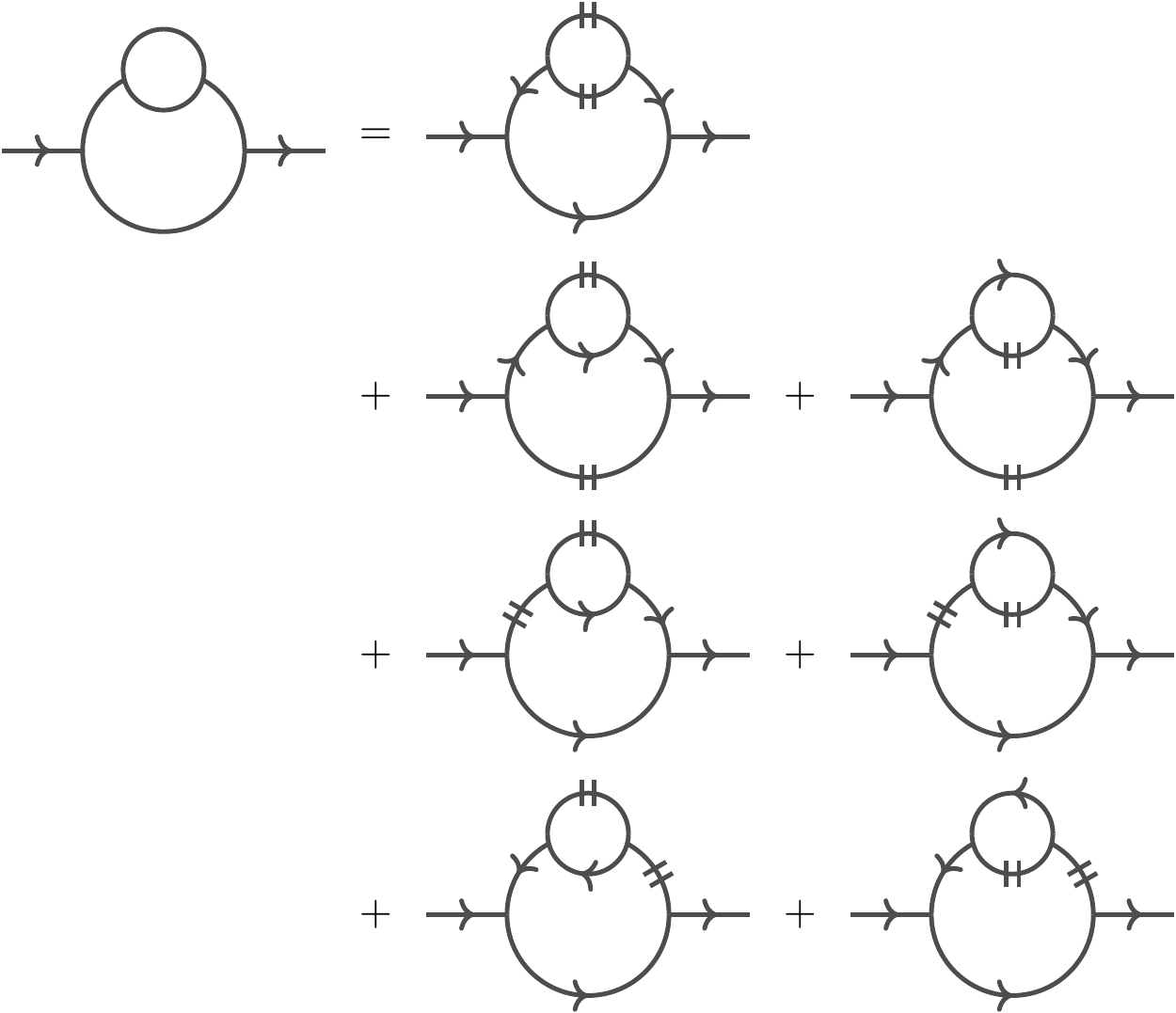}
\end{center}

\begin{center}
    \includegraphics[scale=0.8]{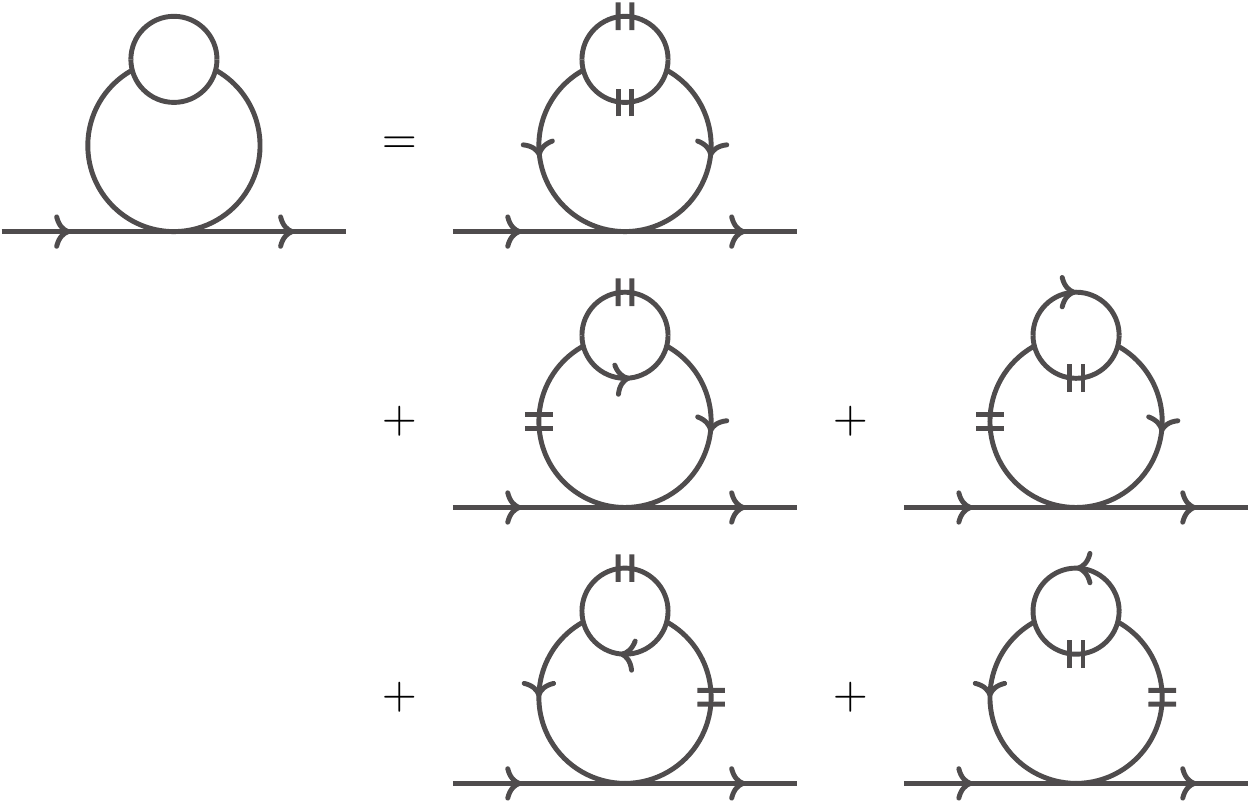}
\end{center}

\begin{center}
    \includegraphics[scale=0.8]{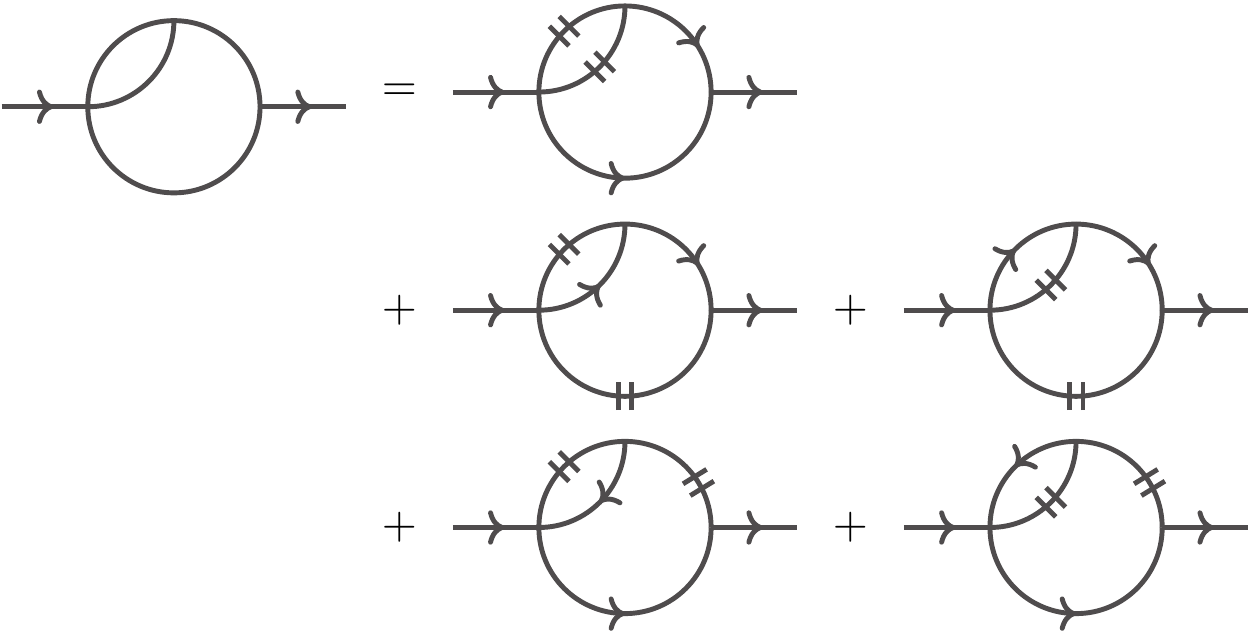}
\end{center}

\begin{center}
    \includegraphics[scale=0.8]{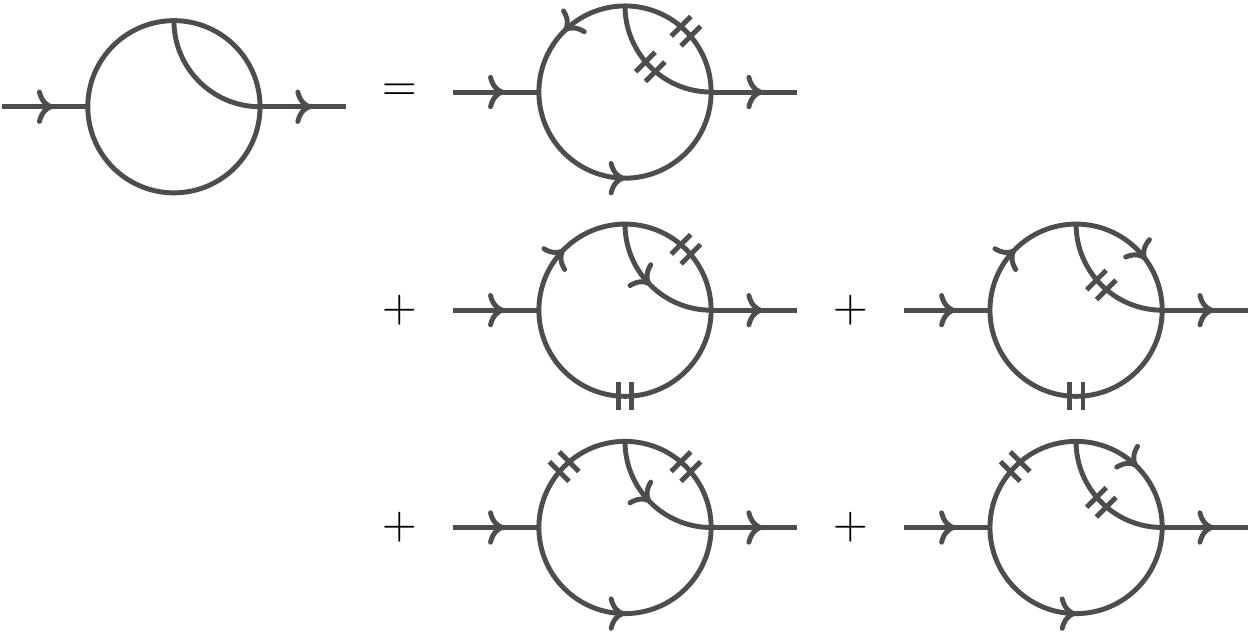}
\end{center}

\begin{center}
    \includegraphics[scale=0.8]{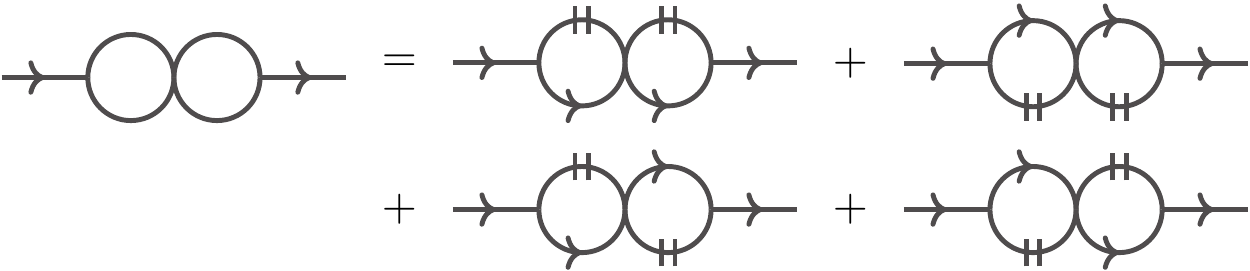}
\end{center}

\begin{center}
    \includegraphics[scale=0.8]{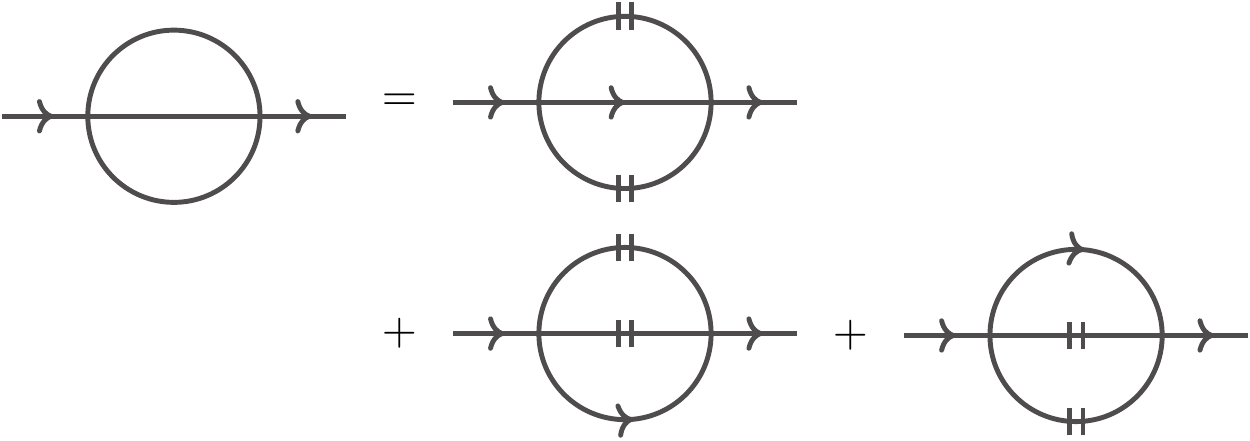}
\end{center}

\begin{center}
    \includegraphics[scale=0.8]{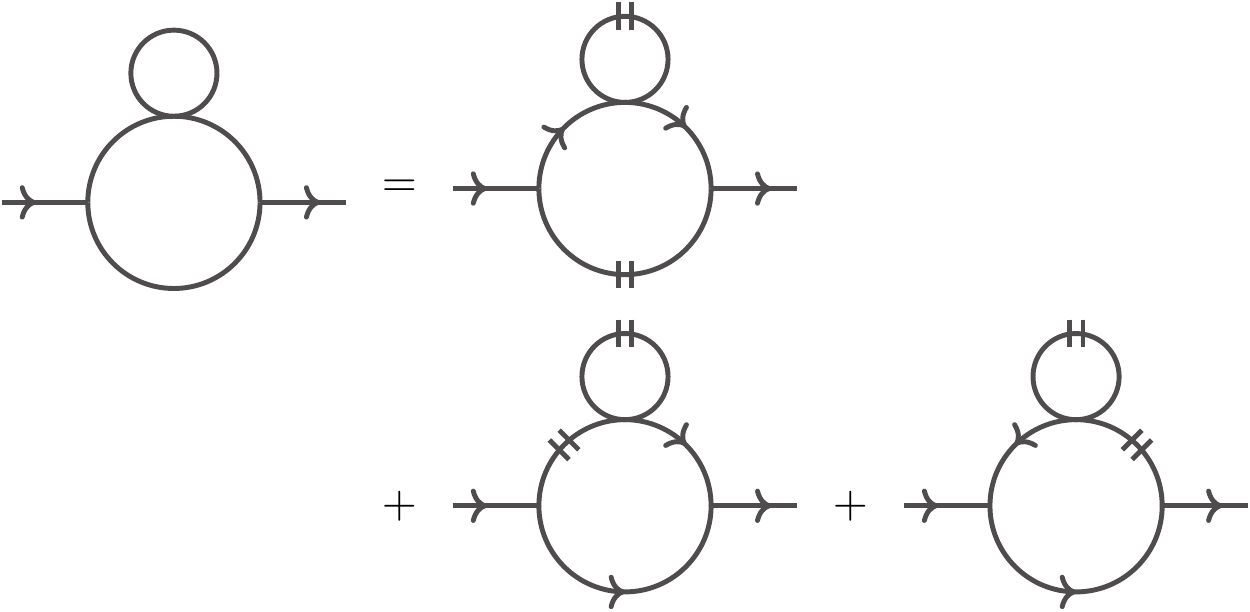}
\end{center}

\begin{center}
    \includegraphics[scale=0.8]{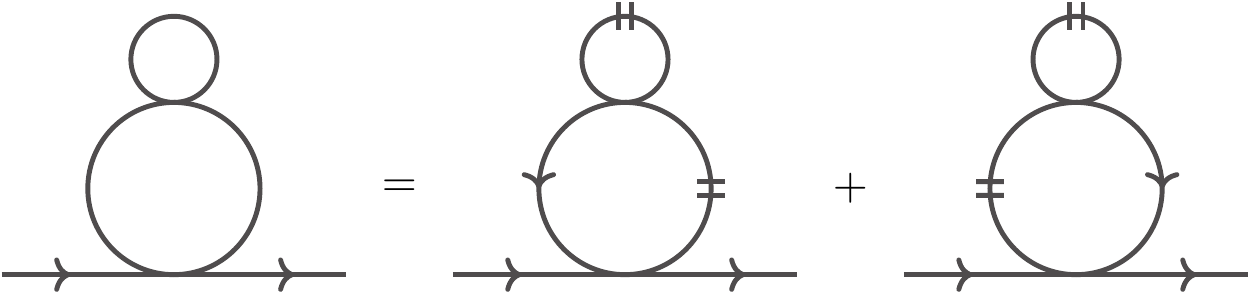}
\end{center}

\section{Gluon contributions to the two-loop self-energy}\label{sec:2loopgluonparts}

With the notation introduced in \cref{sec:twoloopSE} the gluon contributions to the two-loop gluon self-energy read
\begingroup
\allowdisplaybreaks
\begin{align}
\begin{split}
-i\Pi^{\mathrm{2-loop}}_{\mu\nu,\mathrm{gl}_{1}}\delta^{ab} &\equiv -1
    \raisebox{-0.42\height}{\includegraphics[height=1.5cm]{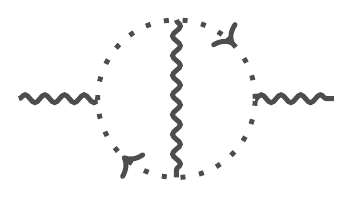}} \\
& = - \sum_{\mathcal{C}}\int_{PQ} \bigl(iV_{\mu}^{gac}(KP)\bigr)\tilde{D}_{c_1}(P)\bigl(iV_{\rho}^{ced}(P)\bigr)\tilde{D}_{c_2}(PQ)\bigl(iV_{\nu}^{dbf}(PQ)\bigr) \\
&\quad\times\tilde{D}_{c_3}(KPQ)\bigl(iV_{\sigma}^{feg}(KPQ)\bigr)\tilde{D}_{c_4}(KP)D^{\rho\sigma}_{c_5}(Q),
\end{split} \\
\begin{split}
-i\Pi^{\mathrm{2-loop}}_{\mu\nu,\mathrm{gl}_{2a}}\delta^{ab} &\equiv -1
    \scalebox{1}[-1]{\raisebox{-0.59\height}{\includegraphics[height=1.5cm]{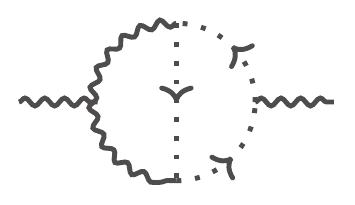}}} \\
& = - \sum_{\mathcal{C}}\int_{PQ} \bigl(iV_{\mu\rho\sigma}^{acd}(-K,-P,KP)\bigr)D^{\rho\lambda}_{c_1}(P)D^{\sigma\gamma}_{c_2}(KP)\bigl(iV_{\lambda}^{gce}(Q)\bigr) \\
&\quad\times\tilde{D}_{c_3}(PQ)\bigl(iV_{\nu}^{ebf}(PQ)\bigr)\tilde{D}_{c_4}(KPQ)\bigl(iV_{\gamma}^{fdg}(KPQ)\bigr)\tilde{D}_{c_5}(Q),
\end{split} \\
\begin{split}
-i\Pi^{\mathrm{2-loop}}_{\mu\nu,\mathrm{gl}_{2b}}\delta^{ab} &\equiv -1
    \scalebox{-1}[1]{\raisebox{-0.42\height}{\includegraphics[height=1.5cm]{figures/SE-2loop/pg11.pdf}}} \\
& = - \sum_{\mathcal{C}}\int_{PQ} \bigl(iV_{\mu}^{eaf}(KPQ)\bigr)\tilde{D}_{c_1}(PQ)\bigl(iV_{\gamma}^{fdg}(PQ)\bigr)\tilde{D}_{c_2}(Q)\bigl(iV_{\lambda}^{gce}(Q)\bigr) \\
&\quad\times\tilde{D}_{c_3}(KPQ)D^{\sigma\gamma}_{c_4}(P)D^{\rho\lambda}_{c_5}(KP)\bigl(iV_{\nu\rho\sigma}^{bcd}(K,-KP,P)\bigr),
\end{split} \\
\begin{split}
-i\Pi^{\mathrm{2-loop}}_{\mu\nu,\mathrm{gl}_{3a}}\delta^{ab} &\equiv -1
    \raisebox{-0.42\height}{\includegraphics[height=1.5cm]{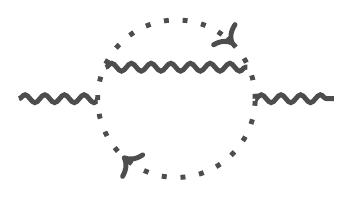}} \\
& = - \sum_{\mathcal{C}}\int_{PQ} \bigl(iV_{\mu}^{dac}(KP)\bigr)\tilde{D}_{c_1}(P)\bigl(iV_{\nu}^{cbg}(P)\bigr)\tilde{D}_{c_2}(KP)\bigl(iV_{\rho}^{gfe}(KP)\bigr) \\
&\quad\times\tilde{D}_{c_3}(KPQ)\bigl(iV_{\sigma}^{efd}(KPQ)\bigr)\tilde{D}_{c_4}(KP)D^{\rho\sigma}_{c_5}(Q),
\end{split} \\
\begin{split}
-i\Pi^{\mathrm{2-loop}}_{\mu\nu,\mathrm{gl}_{3b}}\delta^{ab} &\equiv -1
    \scalebox{-1}[1]{\raisebox{-0.42\height}{\includegraphics[height=1.5cm]{figures/SE-2loop/pg12.pdf}}} \\
& = - \sum_{\mathcal{C}}\int_{PQ} \bigl(iV_{\mu}^{dac}(KP)\bigr)\tilde{D}_{c_1}(P)\bigl(iV_{\sigma}^{cef}(P)\bigr)\tilde{D}_{c_2}(PQ)\bigl(iV_{\rho}^{feh}(PQ)\bigr) \\
&\quad\times\tilde{D}_{c_3}(P)\bigl(iV_{\nu}^{hbd}(P)\bigr)\tilde{D}_{c_4}(KP)D^{\rho\sigma}_{c_5}(Q),
\end{split} \\
\begin{split}
-i\Pi^{\mathrm{2-loop}}_{\mu\nu,\mathrm{gl}_{4}}\delta^{ab} &\equiv -1
    \raisebox{-0.32\height}{\includegraphics[height=1.5cm]{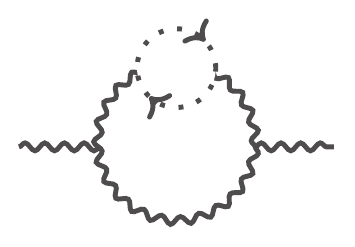}} \\
& = - \sum_{\mathcal{C}}\int_{PQ} \bigl(iV_{\mu\rho\sigma}^{acd}(-K,-P,KP)\bigr)D^{\rho\lambda}_{c_1}(P)\bigl(iV_{\nu\lambda\gamma}^{bce}(K,P,-KP)\bigr) \\
&\quad\times D^{\beta\gamma}_{c_2}(KP)\bigl(iV_{\beta}^{feg}(Q)\bigr)\tilde{D}_{c_3}(KPQ)\bigl(iV_{\alpha}^{gdf}(KPQ)\bigr)\tilde{D}_{c_4}(Q)D^{\alpha\sigma}_{c_5}(KP),
\end{split} \\
\begin{split}
-i\Pi^{\mathrm{2-loop}}_{\mu\nu,\mathrm{gl}_{5}}\delta^{ab} &\equiv -\frac{1}{2}
    \raisebox{-0.42\height}{\includegraphics[height=1.5cm]{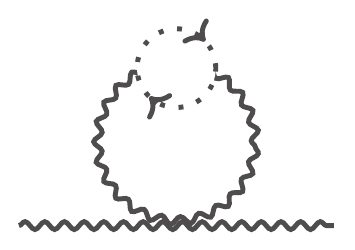}} \\
& = -\frac{1}{2} \sum_{\mathcal{C}}\int_{PQ} \bigl(iV_{\mu\nu\rho\sigma}^{abcd}\bigr) D^{\sigma\gamma}_{c_1}(P) \bigl(iV_{\gamma}^{edf}(PQ)\bigr)\tilde{D}_{c_2}(Q)\bigl(iV_{\lambda}^{fce}(Q)\bigr) \\
&\quad\times\tilde{D}_{c_3}(PQ)D^{\lambda\rho}_{c_4}(P),
\end{split} \\
\begin{split}
-i\Pi^{\mathrm{2-loop}}_{\mu\nu,\mathrm{gl}_{6}}\delta^{ab} &\equiv \frac{1}{2}
    \raisebox{-0.42\height}{\includegraphics[height=1.5cm]{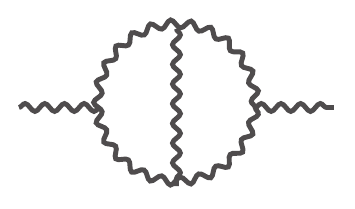}} \\
& = \frac{1}{2} \sum_{\mathcal{C}}\int_{PQ} \bigl(iV_{\mu\rho\sigma}^{acd}(-K,-P,KP)\bigr)D^{\rho\lambda}_{c_1}(P)\bigl(iV_{\lambda\delta\tau}^{cfg}(P,-PQ,Q)\bigr)D^{\delta\alpha}_{c_2}(PQ) \\
&\quad\times\bigl(iV_{\nu\alpha\beta}^{bfe}(K,PQ,-KPQ)\bigr)D^{\beta\eta}_{c_3}(KPQ)\bigl(iV_{\eta\kappa\gamma}^{egd}(KPQ,-Q,-KP)\bigr) \\
&\quad\times D^{\gamma\sigma}_{c_4}(KP)D^{\kappa\tau}_{c_5}(Q),
\end{split} \\
\begin{split}
-i\Pi^{\mathrm{2-loop}}_{\mu\nu,\mathrm{gl}_{7}}\delta^{ab} &\equiv \frac{1}{2}
    \raisebox{-0.42\height}{\includegraphics[height=1.5cm]{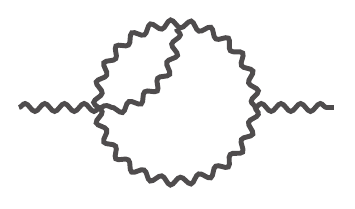}} \\
& = \frac{1}{2} \sum_{\mathcal{C}}\int_{PQ} \bigl(iV_{\mu\rho\sigma\gamma}^{acde}\bigr)D^{\rho\lambda}_{c_1}(P)\bigl(iV_{\nu\alpha\lambda}^{bfc}(K,-KP,P)\bigr)D^{\alpha\beta}_{c_2}(KP) \\
&\quad\times\bigl(iV_{\beta\tau\delta}^{fed}(KP,-KPQ,Q)\bigr)D^{\gamma\tau}_{c_3}(KPQ)D^{\delta\sigma}_{c_4}(Q),
\end{split} \\
\begin{split}
-i\Pi^{\mathrm{2-loop}}_{\mu\nu,\mathrm{gl}_{8}}\delta^{ab} &\equiv \frac{1}{2}
    \raisebox{-0.42\height}{\includegraphics[height=1.5cm]{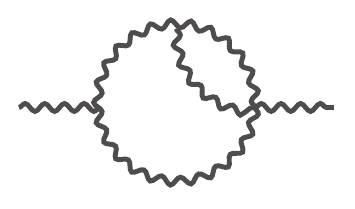}} \\
& = \frac{1}{2} \sum_{\mathcal{C}}\int_{PQ} \bigl(iV_{\mu\rho\sigma}^{acd}(-K,-P,KP)\bigr)D^{\rho\lambda}_{c_1}(P)\bigl(iV_{\nu\alpha\gamma\lambda}^{befc}\bigr)D^{\alpha\beta}_{c_2}(KPQ) \\
&\quad\times\bigl(iV_{\beta\tau\delta}^{edf}(KPQ,-KP,-Q)\bigr)D^{\sigma\tau}_{c_3}(KP)D^{\delta\gamma}_{c_4}(Q),
\end{split} \\
\begin{split}
-i\Pi^{\mathrm{2-loop}}_{\mu\nu,\mathrm{gl}_{9}}\delta^{ab} &\equiv \frac{1}{4}
    \raisebox{-0.42\height}{\includegraphics[height=1.5cm]{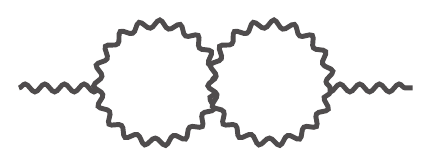}} \\
& = \frac{1}{4} \sum_{\mathcal{C}}\int_{PQ} \bigl(iV_{\mu\rho\sigma}^{acd}(-K,-P,KP)\bigr)D^{\rho\gamma}_{c_1}(P)D^{\sigma\lambda}_{c_2}(KP)\bigl(iV_{\gamma\beta\alpha\lambda}^{cefd}\bigr) \\
&\quad\times D^{\beta\delta}_{c_3}(PQ)D^{\alpha\tau}_{c_4}(KPQ)\bigl(iV_{\nu\tau\delta}^{bfe}(K,-KPQ,PQ)\bigr),
\end{split} \\
\begin{split}
-i\Pi^{\mathrm{2-loop}}_{\mu\nu,\mathrm{gl}_{10}}\delta^{ab} &\equiv \frac{1}{6}
    \raisebox{-0.42\height}{\includegraphics[height=1.5cm]{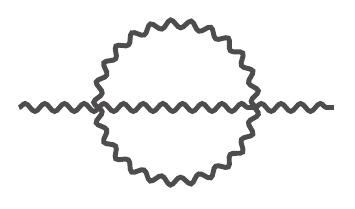}} \\
& = \frac{1}{6} \sum_{\mathcal{C}}\int_{PQ} \bigl(iV_{\mu\rho\sigma\lambda}^{acde}\bigr)D^{\rho\gamma}_{c_1}(P)D^{\sigma\alpha}_{c_2}(KPQ)D^{\lambda\beta}_{c_3}(Q)\bigl(iV_{\nu\beta\alpha\gamma}^{bedc}\bigr),
\end{split} \\
\begin{split}
-i\Pi^{\mathrm{2-loop}}_{\mu\nu,\mathrm{gl}_{11}}\delta^{ab} &\equiv \frac{1}{2}
    \raisebox{-0.33\height}{\includegraphics[height=1.5cm]{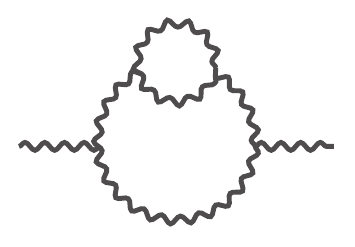}} \\
& = \frac{1}{2} \sum_{\mathcal{C}}\int_{PQ} \bigl(iV_{\mu\rho\sigma}^{acd}(-K,-P,KP)\bigr)D^{\rho\gamma}_{c_1}(P)\bigl(iV_{\nu\lambda\gamma}^{bec}(K,-KP,P)\bigr) \\
&\quad\times D^{\lambda\alpha}_{c_2}(KP)\bigl(iV_{\alpha\delta\eta}^{efg}(KP,-KPQ,Q)\bigr)D^{\delta\tau}_{c_3}(KPQ) \\
&\quad\times\bigl(iV_{\tau\beta\kappa}^{fdg}(KPQ,-KP,-Q)\bigr)D^{\eta\kappa}_{c_4}(Q)D^{\sigma\beta}_{c_5}(KP),
\end{split} \\
\begin{split}
-i\Pi^{\mathrm{2-loop}}_{\mu\nu,\mathrm{gl}_{12}}\delta^{ab} &\equiv \frac{1}{2}
    \raisebox{-0.29\height}{\includegraphics[height=1.5cm]{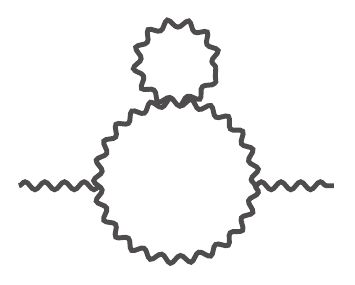}} \\
& = \frac{1}{2} \sum_{\mathcal{C}}\int_{PQ} \bigl(iV_{\mu\rho\sigma}^{acd}(-K,-P,KP)\bigr)D^{\rho\gamma}_{c_1}(P)\bigl(iV_{\nu\lambda\gamma}^{bec}(K,-KP,P)\bigr) \\
&\quad\times D^{\lambda\alpha}_{c_2}(KP)\bigl(iV_{\alpha\tau\delta\beta}^{effd}\bigr)D^{\delta\tau}_{c_3}(Q)D^{\sigma\beta}_{c_4}(KP),
\end{split} \\
\begin{split}
-i\Pi^{\mathrm{2-loop}}_{\mu\nu,\mathrm{gl}_{13}}\delta^{ab} &\equiv \frac{1}{4}
    \raisebox{-0.42\height}{\includegraphics[height=1.5cm]{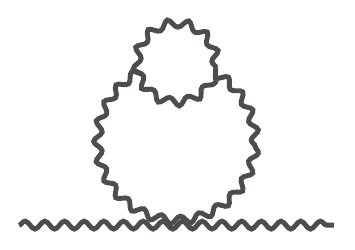}} \\
& = \frac{1}{4} \sum_{\mathcal{C}}\int_{PQ} \bigl(iV_{\mu\nu\rho\sigma}^{abcd}\bigr)D^{\rho\lambda}_{c_1}(P)\bigl(iV_{\lambda\alpha\beta}^{cef}(P,-PQ,Q)\bigr)D^{\tau\alpha}_{c_2}(PQ) \\
&\quad\times\bigl(iV_{\tau\gamma\delta}^{edf}(PQ,-P,-Q)\bigr)D^{\delta\beta}_{c_3}(Q)D^{\sigma\gamma}_{c_4}(P),
\end{split} \\
\begin{split}
-i\Pi^{\mathrm{2-loop}}_{\mu\nu,\mathrm{gl}_{14}}\delta^{ab} &\equiv \frac{1}{4}
    \raisebox{-0.42\height}{\includegraphics[height=1.5cm]{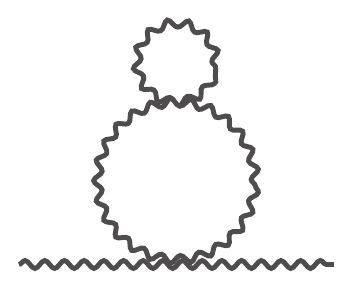}} \\
& = \frac{1}{4} \sum_{\mathcal{C}}\int_{PQ} \bigl(iV_{\mu\nu\rho\sigma}^{abcd}\bigr)D^{\rho\lambda}_{c_1}(P)\bigl(iV_{\lambda\alpha\beta\gamma}^{ceed}\bigr)D^{\alpha\beta}_{c_2}(Q)D^{\sigma\gamma}_{c_3}(P).
\end{split}
\end{align}
\endgroup

\section{Special functions}

We follow \cite{Vuorinen:2003fs} and define 
\begin{equation}
\label{eq:wdef}
w \equiv \frac{1}{2} - i\bar\mu,
\end{equation}
where $\bar\mu\equiv \mu/(2\pi T)$ as in the main text, and define shorthands for some frequently occurring combinations of special functions
\begin{equation}
\label{eq:specialfunctions}
\begin{split}
\aleph(z) & \equiv \Psi(z) + \Psi(z^*), \\
\Psi(z) & \equiv \frac{\Gamma'(z)}{\Gamma(z)}, \\
\aleph(s,z) &\equiv \zeta'(-s,z) + (-1)^s\zeta'(-s,z^*),\\
\zeta'(s,z) &\equiv \partial_s \zeta(s,z),
\end{split}
\end{equation}
where $s,z\in \mathbb{C}$, $\Psi(z)$ is the digamma function, and $\zeta(s,z)$ the Hurwitz zeta function. We find it convenient to express the derivative of the polylogarithm $\mathrm{Li}_{s}(z)$ in terms of the zeta function and its derivatives: 
\begin{equation}
\begin{split}
\mathrm{Li}^{(1)}_{s}(-e^{2\pi \bar \mu}) + \frac{2(-1)^{s}}{1+(-1)^{2s}} \mathrm{Li}^{(1)}_{s}(-e^{-2\pi \bar \mu}) &=\frac{\tan(\pi s)e^{-i\frac{\pi}{2} s}}{(2\pi)^{1-s}}\Gamma(1-s)\Bigg\lbrace  -i \pi e^{i\pi s} \zeta \left(1-s,w^*\right)\\
+\left[\ln(2\pi)+i\frac{\pi}{2} + \pi\cot\left(\pi s\right)-\Psi(1-s)\right]&\left[\zeta(1-s,w)+(-1)^s\zeta(1-s,w^*)\right] -\aleph(s-1,w) \Bigg\rbrace,
\end{split}
\end{equation}
where $\mathrm{Li}^{(1)}_{s}(z) \equiv \frac{\partial }{\partial s}\mathrm{Li}_{s}(z)$, and the relation holds for $\bar \mu \in \mathbb{R}$ and general $s$. 
 
 As we have been unable to find such a relation in the literature, we note that the relation follows for $\operatorname{Re} s>0$ simply by writing the polylogarithm in terms of the zeta function and is readily generalized by analytic continuation. 
 For values of $s$ for which the right-hand side diverges, the relation is taken to hold in a limiting sense; in particular, the pole of the gamma function cancels directly for positive integer $s$.  The nontrivial limit $s \rightarrow 0$ can be taken by using ``Bose-like'' integral representations of the zeta and digamma functions:

\begin{equation}
    \zeta(1-s,z)=\frac{1}{\Gamma(1-s)} \int_{\mathbb{R}_+} \frac{t^{-s}e^{-zt}}{1-e^{-t}} \mathrm{d}t,\quad \Psi(z) = \int_{\mathbb{R}_+} \left(\frac{e^{-t}}{t}-\frac{e^{-zt}}{1-e^{-t}}\right) \mathrm{d}t.
\end{equation}
  
For the particular values of $s$ necessary in the present computation we have 
\begin{equation}
\begin{split}
\mathrm{Li}^{(1)}_{0}(-e^{2\pi \bar \mu}) + \mathrm{Li}^{(1)}_{0}(-e^{-2\pi \bar \mu}) &= - \ln(2\pi e^{\gamE}) - \frac{\aleph(w)}{2}, \\
\mathrm{Li}^{(1)}_{2}(-e^{2\pi \bar \mu}) + \mathrm{Li}^{(1)}_{2}(-e^{-2\pi \bar \mu}) &= -2\pi^2\left[  \ln\left(2\pi e^{\gamE-1}\right)\left(\bar \mu^2 + \frac{1}{12} \right) -\aleph(1,w)\right].
\end{split}
\end{equation}

\bibliography{references}

\end{document}